\documentclass[useAMS,usenatbib]{mn2e}
\usepackage[dvips]{graphicx}
\input{epsf}

\def\la{\mathrel{\mathchoice {\vcenter{\offinterlineskip\halign{\hfil
$\displaystyle##$\hfil\cr<\cr\sim\cr}}}
{\vcenter{\offinterlineskip\halign{\hfil$\textstyle##$\hfil\cr
<\cr\sim\cr}}}
{\vcenter{\offinterlineskip\halign{\hfil$\scriptstyle##$\hfil\cr
<\cr\sim\cr}}}
{\vcenter{\offinterlineskip\halign{\hfil$\scriptscriptstyle##$\hfil\cr
<\cr\sim\cr}}}}}
\def\ga{\mathrel{\mathchoice {\vcenter{\offinterlineskip\halign{\hfil
$\displaystyle##$\hfil\cr>\cr\sim\cr}}}
{\vcenter{\offinterlineskip\halign{\hfil$\textstyle##$\hfil\cr
>\cr\sim\cr}}}
{\vcenter{\offinterlineskip\halign{\hfil$\scriptstyle##$\hfil\cr
>\cr\sim\cr}}}
{\vcenter{\offinterlineskip\halign{\hfil$\scriptscriptstyle##$\hfil\cr
>\cr\sim\cr}}}}}

\title{Limits on the detectability of the CMB B--mode polarization
imposed by foregrounds}

\author[Tucci et al.]{M. Tucci$^1$, E. Mart\'{\i}nez--Gonz\'alez$^1$,
P. Vielva$^{2,3}$ and J. Delabrouille$^{2,3}$ \\
$^1$Instituto de F\'{\i}sica de Cantabria, Consejo Superior de
Investigaciones Cient\'{\i}ficas -- Universidad de Cantabria, \\ Avda. Los
Castros s/n, ~39005 Santander, Spain \\
$^2$F\'ed\'eration de Recherche APC, Universit\'e Paris 7, Paris, France \\
$^3$Physique Corpusculaire et Cosmologie, Coll\`ege de France, 11
pl. M.Berthelot, F-75231 Paris Cedex 5, France}

\pagerange{\pageref{firstpage}--\pageref{lastpage}} \pubyear{2003}

\def\LaTeX{L\kern-.36em\raise.3ex\hbox{a}\kern-.15em
    T\kern-.1667em\lower.7ex\hbox{E}\kern-.125emX}

\begin{document}

\label{firstpage}

\maketitle

\begin{abstract}
We investigate which practical constraints are imposed by foregrounds
to the detection of the B--mode polarization generated by
gravitational waves in the case of experiments of the type currently
being planned. Because the B--mode signal is probably dominated by
foregrounds at all frequencies, the detection of the cosmological
component depends drastically on our ability for removing foregrounds.
We provide an analytical expression to estimate the level of
the residual polarization for Galactic foregrounds, according to the
method employed for their subtraction. We interpret this result in terms
of the lower limit of the
tensor-to-scalar ratio $r$ that allows to disentangle the cosmological
B--mode polarization from the foregrounds contribution. Polarized
emission from extragalactic radio sources and gravitational lensing is
also taken into account. As a first approach, we consider the ideal
limit of an instrumental noise--free experiment: for a full--sky
coverage and a degree resolution, we obtain a limit of
$r\sim10^{-4}$. This value can be improved by high--resolution
experiments and, in principle, no clear fundamental limit on the
detectability of gravitational waves polarization is found. Our
analysis is also applied to planned or hypothetical future
polarization experiments, taking into account expected noise levels.

\end{abstract}


\section{Introduction}
\label{s1}

Standard cosmology is based on the inflationary paradigm, which
provides a good description of the observed Universe. Besides a flat
Universe, inflation predicts a Gaussian, adiabatic and nearly
scale--invariant spectrum of primordial perturbations. However,
inflation is lacking a clear physical support and alternative
theories leading to the present Universe have been investigated (e.g.,
\citealt{kho01}).

A further prediction of inflationary models is the presence of tensor
perturbations to the spatial metric (gravitational waves). The
detection of these perturbations would be an extraordinary way to
distinguish inflation from other competing scenarios and, at the same
time, to constrain the properties of the inflationary potential
\citep{dod97}.

The Cosmic Microwave Background (CMB) anisotropies and polarization
probably offers the best means (if not the only) of detecting the
tensor metric perturbations. In fact, in most inflationary models, the
amplitude of the power spectrum for ``tensor'' temperature
anisotropies is directly related to the energy scale at which the
inflation has occurred. Considering a single--field inflation (to
lowest order in the slow--roll parameters), the relation between the
inflationary potential, $V(\phi)$, and the tensor quadrupole $Q_T^2$
is \citep{tur96}
\begin{equation}
{V_{\ast} \over m_{Pl}^4}\simeq1.65\,Q_T^2\,,
\label{e1}
\end{equation}
where $V_{\ast}$ is the inflationary potential evaluated when the
present Hubble scale crossed outside the horizon during inflation.
Then, the ``energy scale'' of the inflation $E_i=V_{\ast}^{1/4}$ can
be expressed in terms of the scalar quadrupole $Q_S^2$ and the ratio
of tensor to scalar quadrupoles $r=Q_T^2/Q_S^2$\footnote{The
tensor-to-scalar ratio is often defined also in terms of the
primordial amplitude of tensor and scalar fluctuations
\citep{lea02}. For the ``concordance'' model, it corresponds to
$1.78\,Q_T^2/Q_S^2$.}
as
\begin{equation}
E_i\simeq Q_S^{1/2}r^{1/4}m_{Pl}\simeq3\times10^{16}\,r^{1/4}\,{\rm GeV}\,,
\label{e1a}
\end{equation}
where we have used the COBE estimate $Q_S\simeq18\mu K$ \citep{hin96}.
However, tensor perturbations contribute to the temperature power
spectrum essentially at low $\ell$, with a roughly scale--invariant
spectrum. Unless the relative amplitude of tensor perturbations
respect to scalar ones is large enough ($r>0.1$), measurements of the
temperature power spectrum can only provide an upper limit on $Q_T^2$
due to constraints imposed by the cosmic variance \citep{kno94} [for
example, WMAP finds $r<0.71$ \citep{spe03}].

The detectability of the contribution of tensor perturbations is much
more promising exploiting CMB polarization measurements. The
polarization field can be divided into a curl--free component of even
parity called ``E--mode'', and a curl component of odd parity,
``B--mode'' \citep{kam97,zal97}. While the former is generated by both
scalar and tensor perturbations, B--mode polarization arises only from
tensor perturbations and carries thus a direct imprint of the
inflationary epoch. Nevertheless, the B--mode intensity is expected to
be extremely weak, and, even for optimistic values of $r$, the rms
signal is only a fraction of $\mu$K, less than 1$\%$ of the level of
temperature anisotropies at degree scales. Future experiments will
need to reach extraordinary sensitivities to be able to detect such a
low signal.

Gravitational waves produced during inflation are not the only
possible source of primordial B--mode polarization. For example, the
existence of tangled primordial magnetic fields or cosmic strings
would generate a vector component of metric perturbations, providing a
contribution to the B--mode polarization on small angular scales
\citep{ses01,sub03,pog03}. In any case, the discussion of these
sources of polarization is outside the scope of this work.

Even if the instrumental sensitivity appears to be the biggest problem
for detecting the tensor--induced polarization, important constraints
come also from the presence of other sources of the B--mode
polarization, as foregrounds and effects mixing CMB E-- and B--modes.
The most relevant mode--mixing effect is due to the gravitational
lensing produced by large scale structures, which converts a fraction
of the CMB E--mode component to B--mode \citep{zal98}. Different
authors (\citealt{hu02}, \citealt{kes03},
\citealt{sel03}) presented methods to reconstruct the gravitational
lensing potential using information from the CMB polarization itself
and to remove the lensing contamination. The problem of a
correct separation between E-- and B--mode in the presence of a
partial sky coverage, pixelization and systematic effects has been
also widely dealt with in literature (see \citealt{lew02},
\citealt{bun02}, \citealt{bun03}, \citealt{hu03}, \citealt{lew03},
\citealt{cha04}), and a suitable formalism to minimize their influence
has been provided.

Another critical source of confusion for the detection of the CMB
polarization is expected from foreground contamination. Apart from the
free--free emission that is not polarized, foregrounds have been shown
to have, in general, an high degree of polarization equally shared
between E-- and B--mode: synchrotron emission is on average
$20\%$--$30\%$ polarized while dust and extragalactic sources are
polarized from few per cents to 10 per cent or more. Foregrounds are
expected to dominate the sky B--mode polarization at all frequencies
and angular scales, even for high values of $r$. On the other hand,
they are also a serious problem for the CMB E--mode signal which is at
most a 10$\%$ of temperature anisotropies, but significantly less on
large angular scales.

A secondary B--mode polarization is also generated by scattering of
the CMB photons from ionized gas in galaxy clusters or in a patchy
reionization scenario (see \citealt{hu00}, \citealt{liu01},
\citealt{bau03}, \citealt{san03}). However, these effects produce a
polarization that is several orders of magnitude below the dominant
contributions.

The main question that this paper address is how Galactic and
extragalactic foregrounds (including the gravitational lensing
contamination) limit the detectability of the CMB B--mode
polarization. In terms of cosmological constraints, we want to
estimate the lower limit of $r$ (hence of the energy scale of
inflation) under which the signal from tensor perturbations cannot be
distinguished by polarization measurements. The case of an ideal
experiment (i.e., noise--free experiment) is first considered. Then,
the same analysis is also applied to planned or hypothetical
experiments, giving hints on the best strategy for future missions.

\section{Intensity level of B--mode polarization for CMB and foregrounds}
\label{s2}

\begin{figure*}
\includegraphics[width=84mm]{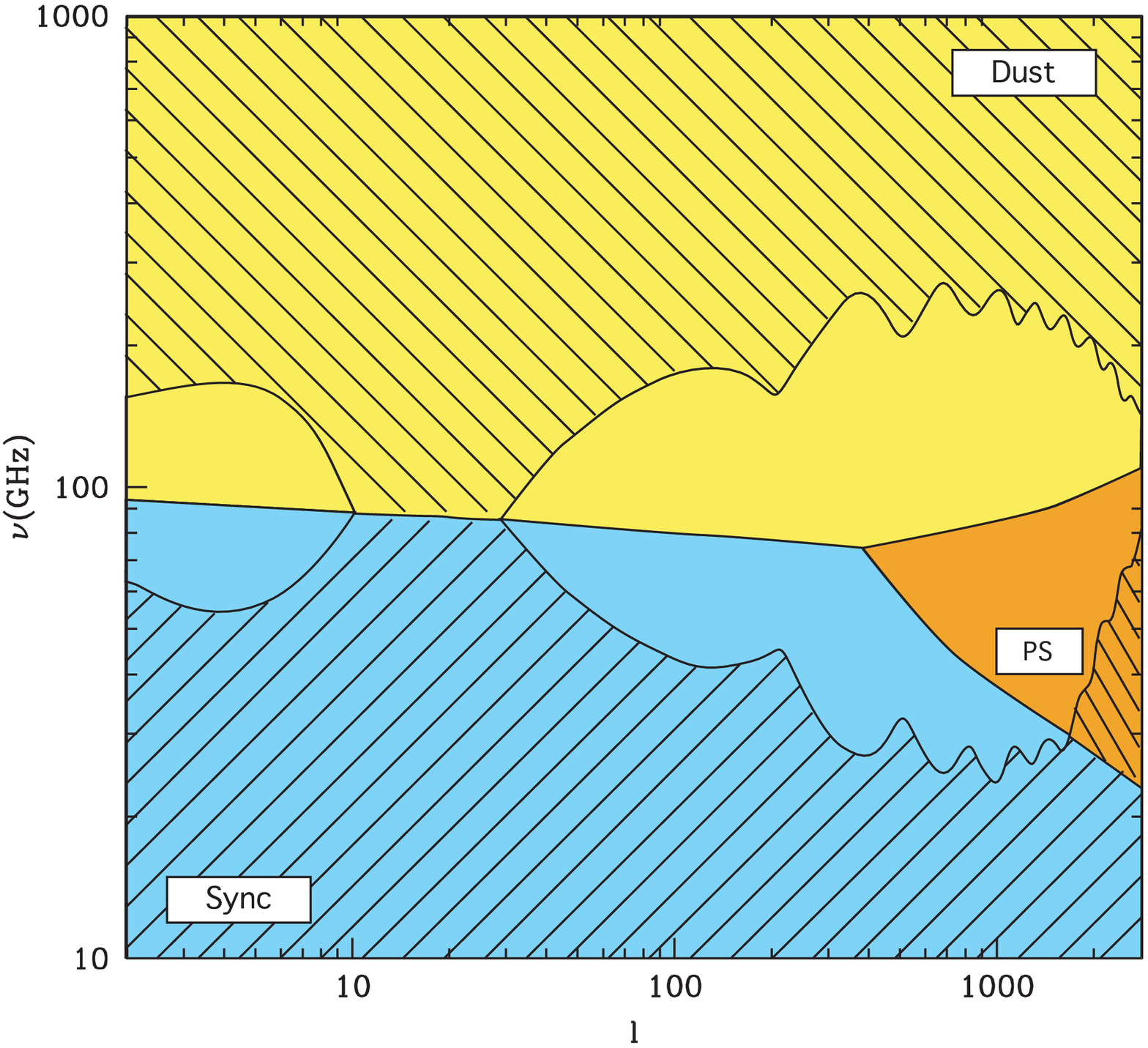}
\includegraphics[width=84mm]{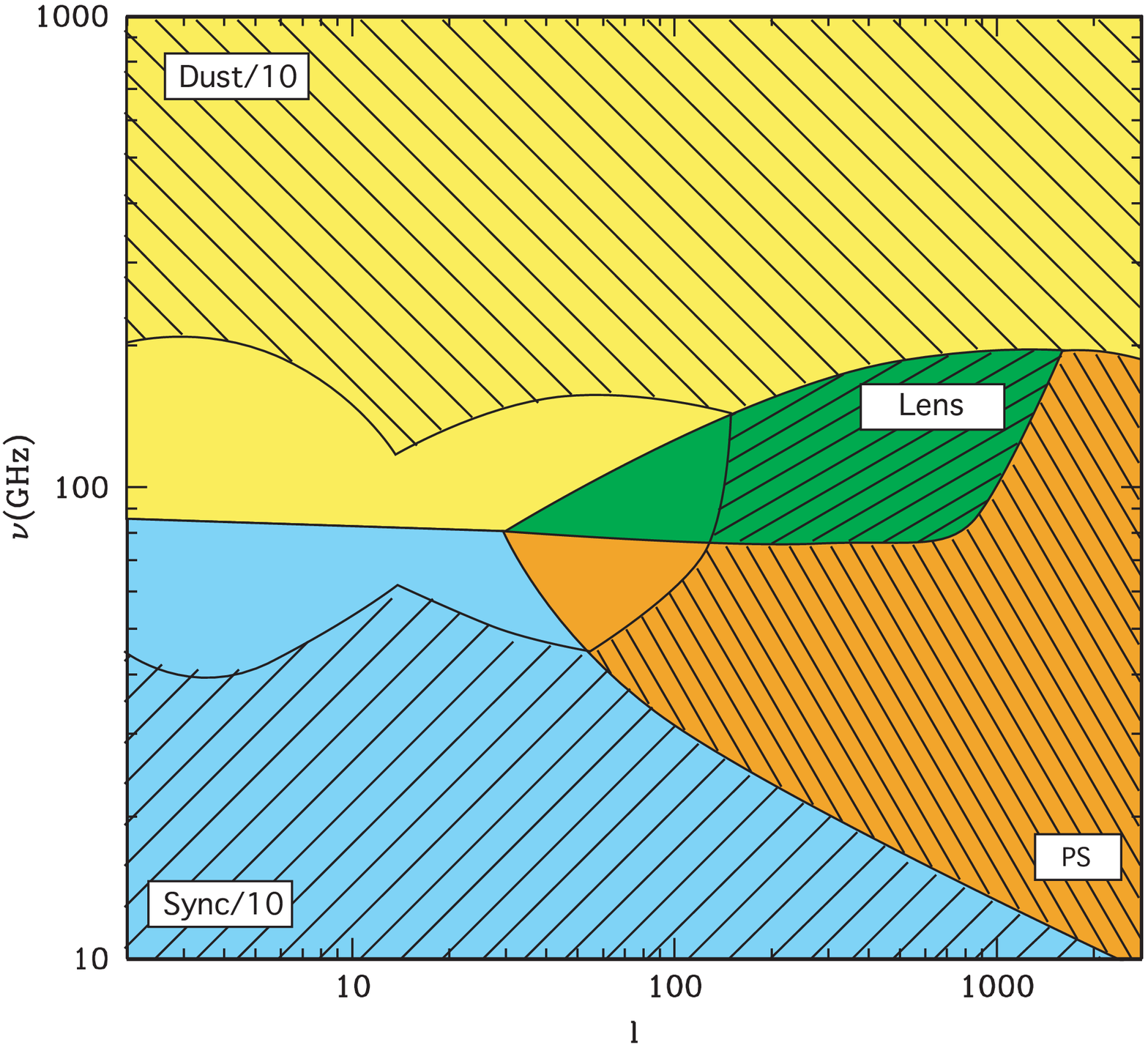}
\caption{The dominant foreground in polarization as function of the
frequency and the angular scale (synchrotron: light blue; dust: yellow;
extragalactic radio sources: brown; gravitational lensing: green).
Galactic foregrounds are modelled by Eq. (\ref{e16a})--(\ref{e16b})
(see also Figure 2b).
For extragalactic radio sources we use a flux limit of 1\,Jy.
{\bf Left plot}: the not--shaded area is where the CMB E--mode is
the dominant polarized signal. {\bf Right plot}: as in the left plot
but after reducing by a factor 10 the synchrotron and dust emission.
Now, the not--shaded area indicates the region where the CMB B--mode
(considering $r=0.1$) is higher than the foregrounds contribution.}
\label{f0}
\end{figure*}

\begin{figure*}
\includegraphics[width=84mm]{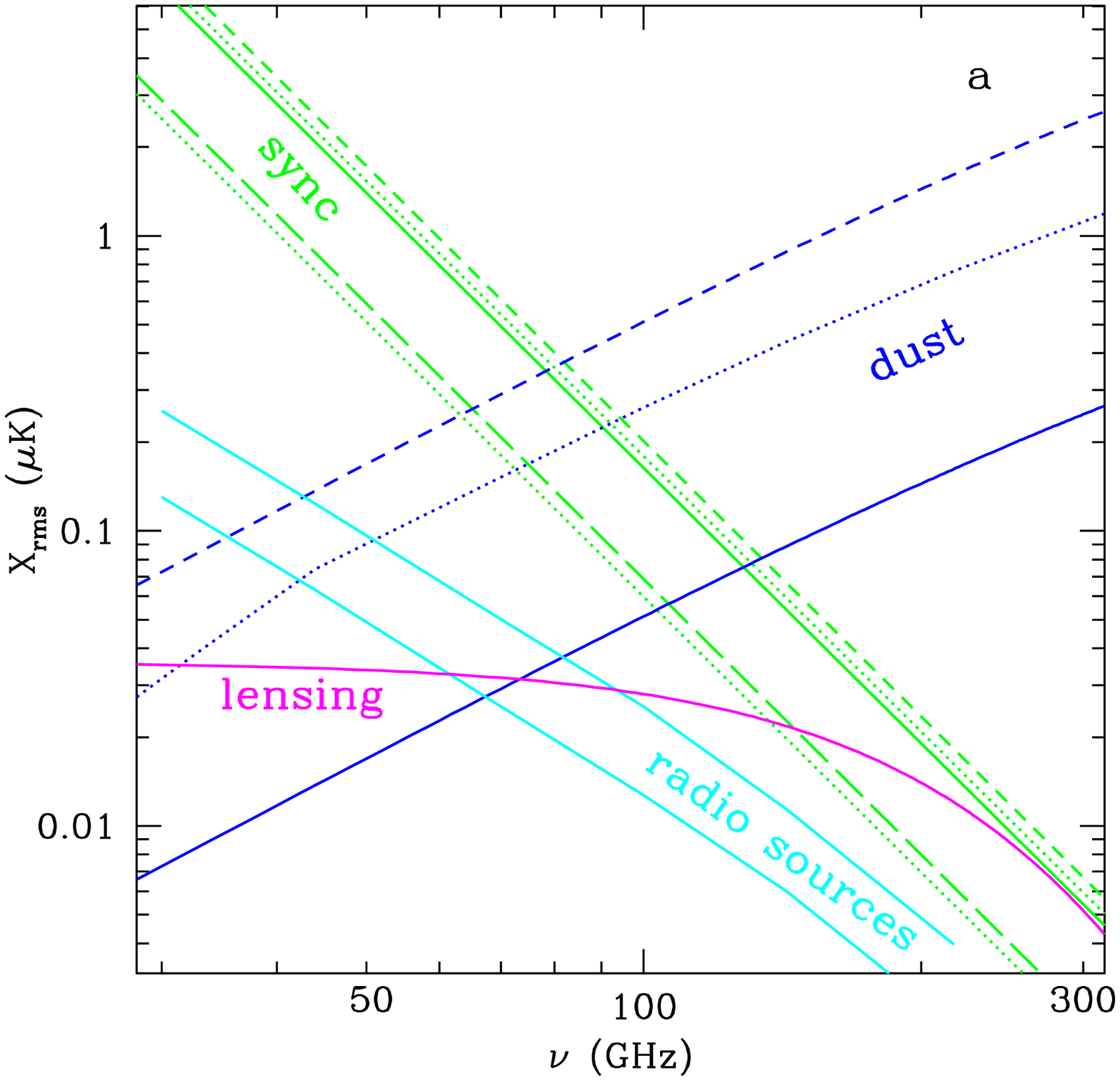}
\includegraphics[width=84mm]{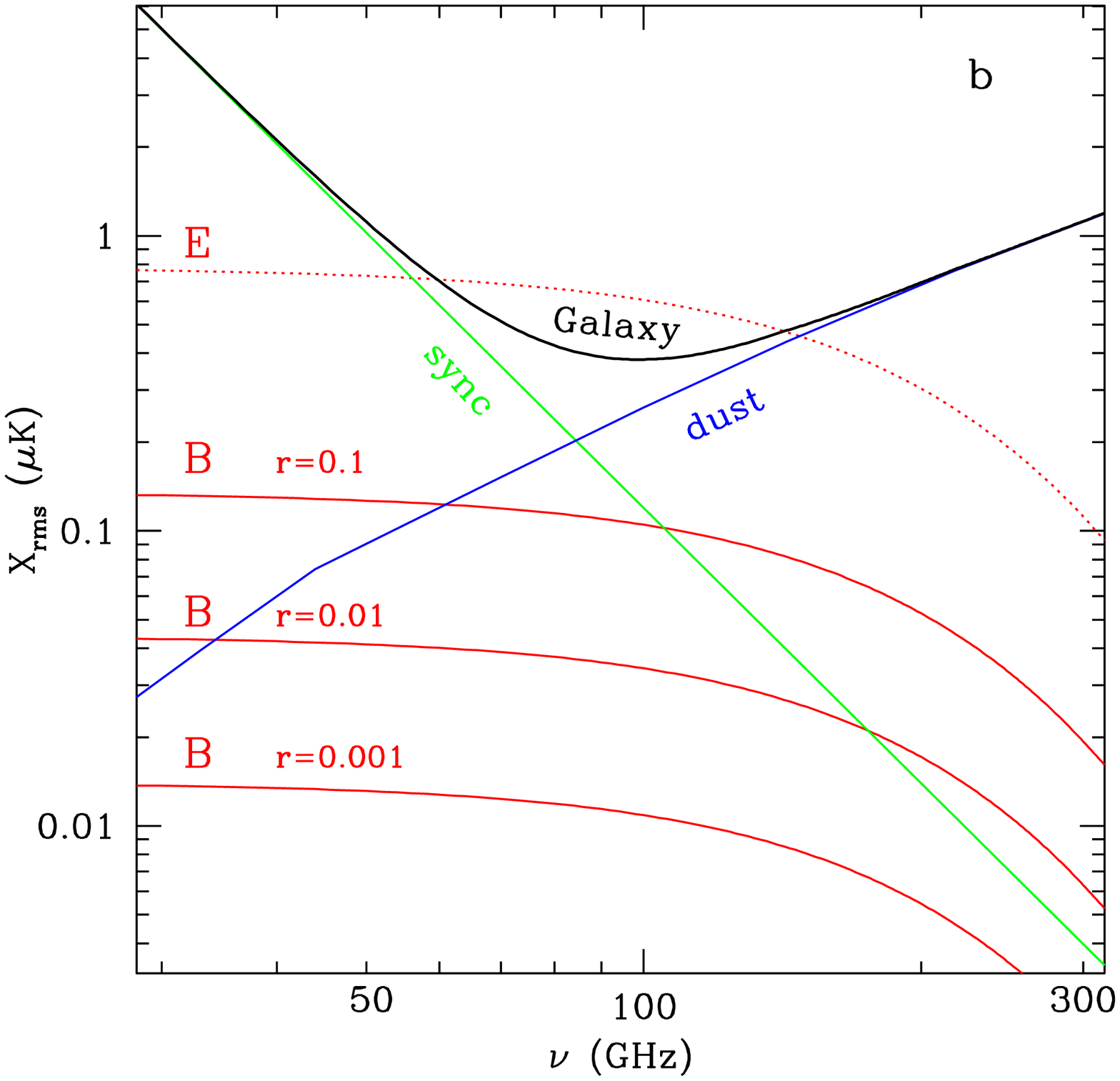}
\caption{The rms value of foreground and CMB X=E,\,B modes vs. the
frequency, as expected by an experiment with FWHM$\,=1\degr$. \newline
--{\bf a}-- Estimates of foregrounds from different data.
{\bf Synchrotron (green lines)}: 10$\%$--30$\%$ of the WMAP $\Delta
T_{\rm rms}$ (dotted lines); estimates from Brouwn\,\&\,Spoelstra
data (dashed line), from high--resolution low--latitude polarization
surveys (solid line) and from observations of high--latitude
polarization (\citealt{abi03}; long--dashed line). The spectral index
is $\beta_s=-3.1$. {\bf Dust (dark
blue lines)}: 5$\%$ of the WMAP $\Delta T_{\rm rms}$ (dashed line)
and the result from the \citet{pru98} model (solid line). For the
frequency spectrum we consider a one--component dust model with
$T=18\,K$ and emissivity $\alpha=1.7$. The dotted line is obtained by
the ``100$\mu$m map'' \citep{sch98} extrapolated to microwave
frequencies using the model 8 of \citet{fin99}. {\bf Radio sources
(light blue lines)}: estimates from \citet{tuc04}, taking the flux
limit 1 and 0.2\,Jy. {\bf Lensing--induced polarization (magenta
line)}. \newline
--{\bf b}-- The black solid line is the total $X_{rms}$ contribution
of foregrounds, taking for the synchrotron the 20$\%$ of the WMAP
estimate and $\beta_s=-3.1$; for the dust the model 8 of \citet{fin99}.
{\bf CMB (red lines)}: $E_{rms}$ (dotted line) and $B_{rms}$ (with
$r=0.1,\,0.01,\,0.001$; solid lines) for the ``concordance'' model
with $\tau=0.1$.}
\label{f1}
\end{figure*}

In this section we discuss the expected level of the polarized
signal for the CMB and foreground emission as a function of the
observing frequency $\nu$ and the angular scale. The results are
presented in Figures \ref{f0} and \ref{f1}. The left plot of Figure
\ref{f0} shows the dominant foregrounds in the $\ell$--$\nu$
space. The not--shaded area is where the CMB E--mode is higher than
the foregrounds E--mode contribution. The diffuse Galactic emission
(synchrotron and dust emission) is essentially the strongest
foreground in polarization. Only at high $\ell$ and frequencies lower
than 100\,GHz extragalactic radio sources start to be relevant. We
notice how the ``cosmological window'' (the $\ell$--$\nu$ region where
CMB exceeds foregrounds emission) is strongly reduced as compared to
temperature fluctuations, expecially at large angular scales. In the
right plot, we consider the polarization level of Galactic emissions
but reduced by an order of magnitude (this is, at least, what we
expect to reduce in a multyfrequency experiment). In this case,
extragalactic radio sources become an important contribution already
for $\ell\ga100$, as well as the polarization induced by gravitational
lensing at $100\la\nu\la200$\,GHz. Now the not--shaded area represents the
$\ell$--$\nu$ region dominated by the CMB B--mode polarization for a
cosmological model with $r=0.1$. As expected, it is confined only to
$\ell\la100$.

In Figure \ref{f1} we plot the rms value for the $X=E,\,B$ modes
polarization\footnote{Hereafter, we are supposing that, on average,
the amplitude of E-- and B--mode polarization is the same for
foregrounds emission.} as expected to be measured by an experiment
with a resolution of $1\degr$:
\begin{equation}
X_{\rm rms}=\Big[{T^2_0 \over 4\pi}\sum(2\ell+1)C_{X\ell}W^2_{\ell}
\Big]^{1/2}\,,
\label{e1b}
\end{equation}
where $W_{\ell}$ is the beam function of a hypothetical instrument
and $C_{X\ell}$ is the angular power spectrum (APS) of CMB or
foregrounds. The rms values are given in antenna temperature. The
$1\degr$ resolution is chosen to estimate the rms polarization
because degree angular scales are the most interesting for detecting
the CMB B--mode polarization. Let us see now the properties for each
polarized component in detail.

\subsection{Cosmic microwave background}
\label{s2a}

The CMB B--mode power spectrum ($C_{B\ell}$) is characterized by a
peak at $\ell_p\sim90$, i.e. at the angular scale corresponding
to the horizon at recombination. The amplitude of the spectrum is
directly proportional to the cosmological parameter $r$ and is related
to the energy scale of inflation: at the peak the amplitude is
\begin{eqnarray}
\Delta B_p & \equiv & \bigg[{\ell_p(\ell_p+1) \over 2\pi}
C_{B\ell_p}\bigg]^{1/2}T_0\sim \nonumber \\
& \sim & 0.3\,r^{1/2}\,\mu{\rm K}
\simeq0.03\bigg({E_i \over 10^{16}{\rm GeV}}\bigg)^2\,\mu{\rm K}\,.
\label{e2}
\end{eqnarray}
At $\ell>\ell_p$ the spectrum rapidly decreases because gravitational
waves oscillate and decay once inside the horizon. At very large
scales very low B--mode polarization is also expected because, at the
moment of recombination, the anisotropy quadrupole moment has not been
significantly produced yet on scales larger than the horizon. However,
the reionization of the Universe has the effect to partially polarize
the CMB radiation and shift to lower redshift a fraction of the last
scattering, producing a further peak at low $\ell$ in the spectrum.
The amplitude and the position of such a peak depends on the
optical depth of the Universe, $\tau$ [WMAP finds $\tau=0.17\pm0.07$
\citep{spe03}].

In Figure 2b we report the expected values of $B_{\rm rms}$ for
$r=0.1,\,10^{-2}$ and $10^{-3}$ (solid red lines).
As shown in Eq. (\ref{e2}), it scales like $r^{1/2}$. If only very
large scales are considered, $B_{\rm rms}$ is strongly dependent on
$\tau$ and exceeds 0.1\,$\mu$K for $\tau$ and $r>0.1$. On
the contrary, at degree resolution, $\tau$ is less relevant but not
negligible ($B_{\rm rms}$ increases by a factor 2 changing $\tau$
from 0 to 0.17). In the plot we consider $\tau=0.1$ (a conservative
value compared to the WMAP best--fit one). The B--mode
spectrum is less sensitive to other cosmological parameters. We fix
those in agreement with the ``concordance model'' ($\Omega_0=1$,
$\Omega_{\Lambda}=0.7$, $\Omega_b=0.05$ and $h=0.7$).

The detection of the CMB E--mode polarization is much easier than that
of B--mode and it is even better with sub--degree resolution
experiments (see Figure \ref{f0}).
At these angular scales, DASI \citep{kov02} achieved the first direct
measure of E--mode polarization, while more recent observations have
been provided by DASI \citep{lei04}, CAPMAP \citep{bar05} and CBI
\citep{rea04}.
At degree angular scales the value
of $E_{\rm rms}$ is less than $1\mu$K with a polarization degree of
only few per cent or less. Nevertheless, the CMB E--mode is still the
dominant contribution in a wide range of frequencies, approximately
between $60<\nu<150$\,GHz.

Finally, we plot in Figure 2a the level of the B--mode polarization
induced by the gravitational weak lensing of large scale structures
(see magenta line). This estimate is obtained using the CMBFAST
package\footnote{http://www.cmbfast.org}.

\subsection{Synchrotron emission}
\label{s2b}

Among all foregrounds, synchrotron emission is the most polarized (up
to $\sim75\%$). It is expected to be the dominant component at low
frequencies ($\nu\la70\,$GHz). Total--intensity data are available up
to $\sim20$\,GHz, while polarization observations of the synchrotron
are obtained only at few GHz or less. Estimates on the amplitude of
the polarized synchrotron emission at cosmological frequencies are
especially complicated for two reasons: the steepening of the spectral
index that may occur at $\nu\ga20$\,GHz, and the frequency dependence
of the degree of polarization.

The synchrotron spectral index, $\beta_s$, is computed on nearly all
sky using the surveys at 408\,MHz \citep{has82}, at 1.4\,GHz
\citep{rei86} and at 2.3\,GHz \citep{jon98} by \citet{gia02} and
\citet{pla03}. These analysis show $\beta_s$ to vary across the
sky between $-2.5$ and $-3.0$, with an average value $<\beta_s>
\simeq-2.7$ and a dispersion of 0.1--0.2 at sub--degree angular
resolution. A relevant steepening ($<\beta_s>\approx-3$ out of the
Galactic Plane) is found from the first--year data of WMAP at 23\,GHz
\citep{ben03}. \citet{ber04} studied the distribution of the spectral
index computed between 1.4 and 23\,GHz: the frequency spectrum is
observed rather flat along the Galactic Plane, whereas steeper at high
Galactic latitudes with a typical value $\beta_s=-3.1$. A further
steepening is also suggested in the 23--41\,GHz range by \citet{ben03},
although it has to be confirmed. In the following, we shall assume
$\beta_s=-3.1$ as the mean spectral index for the synchrotron emission
at all considered frequencies.

Estimates of the rms signal of $X=E,\,B$ modes for the synchrotron
emission are reported in Figure 2a (green lines). A set of
estimates obtained in different ways is considered, which permits to
get a range of plausible levels. First we scale down unpolarized
synchotron as follow: the full--sky synchrotron map at 23\,GHz,
obtained from WMAP data after component separation with the Maximum
Entropy Method \citep{ben03} and smoothed by a Gaussian beam with
FWHM$=1{\degr}$, is used to compute the rms amplitude of temperature
fluctuations at high latitudes ($|b|>20\degr$). We find $\Delta
T_{\rm rms}=82.6$. From the \citet{bro76} observations (the only
up--to--now available data of the synchrotron polarization on a large
fraction of the sky), \citet{spo84} found that on degree scales the
fractional polarization at 1.411\,GHz is typically 10--20 per cent
with the highest value being 35$\%$. Taking into account that Faraday
rotation can reduce the polarization degree at this frequency, we
assume that synchrotron emission is polarized between a 10 and a 30
per cent as lower or upper limit. We get estimates as plotted with
dotted lines in Figure 2a.

Secondly, estimates of $X_{\rm rms}$ can be directly achieved also
from polarization data. \citet{bru02} use \citet{bro76} survey to
compute the E-- and B--mode APS in three patches with high
signal--to--noise ratio centered at the latitudes
$b=5{\degr},\,44.5{\degr},\,72.5{\degr}$. They find that the APS can
be well approximated by a power law with a quite flat slope, $\ga-2$.
Taking as the ``average'' spectrum at 1.4\,GHz 
$C_{X\ell}=0.01\ell^{-1.8}$(K$^2$), we find 
$X_{\rm rms}(1.4{\rm GHz})=0.11$K at $1\degr$ resolution (dashed line
in Figure 2a). Similar results are also obtained from the
analysis of high--resolution surveys at low Galactic latitudes
(\citealt{dun97}, \citealt{dun99}, \citealt{uya99}): extrapolating the
polarization APS provided by \citet{tuc00} and \citet{bru02} to low
$\ell$s, we find $X_{\rm rms}\simeq0.02$\,K at 2.4\,GHz (solid line in
Figure 2a). However, these estimates probably have to be
considered as upper limits because they come from observations of the
Galactic Plane or of highly--polarized regions. Their very good
agreement with the 30$\%$--WMAP predictions shown by Figure 2a
confirms that a polarization degree up to 30$\%$ can be observed in
highly--polarized area but as it may be not the typical percentage on
the sky. The upper lines in Figure 2a may be somewhat pessimistic
estimates.

Finally, the recent observations of the high--Galactic latitude
polarization with the Effelsberg Telescope at 1.4\,GHz \citep{abi03}
point out structures in polarization on scales of several degrees with
a fractional polarization up to 30--40$\%$ in the brightest regions
and a rms $Q$ and $U$ signal of 50\,mK. This value, extrapolated to
WMAP frequencies, corresponds to the WMAP result with an average
polarization degree of 10$\%$ (see long--dashed line). Nevertheless,
as the polarization level at 1.4\,GHz can be significantly reduced by
Faraday depolarization, a percentage of 20$\%$ for the polarized
synchrotron emission may be more realistic at microwave frequencies.
The lower lines in Figure 2a are thus probably somewhat optimistic.

\subsection{Dust emission}
\label{s2d}

Millimetric or submillimetric measurements of the Galactic dust
polarization are usually concentrated in Galactic clouds and
star--forming regions with arcminutes angular resolution (e.g,
\citealt{hil99}, \citealt{gre99}). They show a few per cent
polarization and no clear frequency dependence. The first
observations on large angular scales of the polarization of the
Galactic dust emission are provided by Archeops at 353\,GHz
\citep{beno03}. The Archeops data show a significantly large--scale
polarized emission in the Galactic Plane, with a polarization degree
of 4--5$\%$, with several clouds of few square degrees appearing to
be polarized at more than 10$\%$.

Total--intensity observations at infrared wavelengths, where the
thermal dust emission is very dominant, have been obtained by the
IRAS and COBE (DIRBE) satellites. Combining these data, \citet{sch98}
generated a full--sky map at 100$\mu$m with sub--degree resolution. A
tentative extrapolation of the 100$\mu$m map to microwave frequencies
was done by \citet{fin99}, assuming different dust emissivity models.
They use the FIRAS data in the 100--2100\,GHz range to constrain dust
properties. The best agreeement is met with a two--component model,
consisting of a mixture of silicate and carbon--dominated grains (see
Model 8 of \citealt{fin99}). Extrapolating the 100$\mu$m map to
100\,GHz with this model, we find $\Delta T_{\rm rms}=7.4\,\mu$K for a
$1\degr$ resolution. The dust rms intensity at this resolution can
also be estimated from WMAP data: using the Galactic emission models
by \citet{ben03}, it is found $\Delta T_{\rm rms}=13\,\mu$K at
93.5\,GHz, significantly higher than the \citet{fin99} value. However,
because in the WMAP frequency range the dust emission is only a
secondary contribution, this value is still controversial.
The predictions on the rms polarization presented in Figure 2a
assume a typical polarization degree for the interstellar diffuse
dust emission of 5$\%$, in agreement with Archeops results (dotted
and dashed blue lines).

Finally, \citet{pru98} provide a model of the dust polarized emission
using the three--dimensional HI maps of the Leiden--Dwingeloo survey
at high Galactic latitudes. Different hypothesis on dust grain
properties and alignment are assumed. As main result, they estimate
the dust polarization APS at 100\,GHz to be
$C_{X\ell}\simeq10^{-3}\ell^{-1.4}\,(\mu$K)$^2$, corresponding to
$X_{\rm rms}=0.051\,\mu$K at $1\degr$ resolution (solid line in Figure
2a). This value is significantly low compared to previous ones and
require a mean polarization degree less than $1\%$ for being
consistent with total--intensity results.

\subsection{Extragalactic sources}
\label{s2c}

Extragalactic radio sources contribute significantly to
temperature and polarization fluctuations only at small angular
scales. At degree scales their polarized signal is a small
fraction of the total foregrounds, even if it can exceed the
cosmological B--mode for a low value of $r$. Predictions on the
polarization properties of radio sources at microwave frequencies and
on their contamination for CMB measurements are provided by \citet{tuc04},
using the evolutionary model of \citet{tof98} which predicts the
number counts of sources at cm and mm wavelengths. In Figure 2a we
report the rms polarization assuming that all sources with flux
density higher than 1 or 0.2\,Jy have been completely removed (light
blue lines).

At frequencies higher than 100\,GHz, the contribution of dusty
galaxies should also be taken into account. However, the sparsity
of data on polarization at sub-mm wavelengths makes reliable
predictions difficult. In any case, we expect that the
polarization degree from the dust emission in external galaxies is not
higher than the polarization degree observed in the Milky Way,
i.e. only a few per cent. Supposing a polarization of 2$\%$
\citep{hil96,gre00,mat02}, Figure \ref{f0} shows that the dusty
galaxies contribution never overcomes the other foregrounds at all
frequencies lower than 1000\,GHz and at $\ell<3000$. Polarization from
dusty galaxies will be neglected in this work.

\subsection{Other sources of polarization}
\label{s2d}

Here we give a brief description of some secondary sources of
polarized emission, although they will not be taken into account in
the analysis.

The existence of an anomalous dust--correlated emission, far brighter
than the expected thermal dust emission, has been well detected at
$\sim10$--30\,GHz (e.g., \citealt{kog96}, \citealt{deo99},
\citealt{fin03}). One possible explanation is the electric dipole
emission from rapidly rotanting small dust grains, known as ``spinning
dust'' \citep{dra98}. This emission is expected to be polarized if
grains are aligned with the magnetic field. In any case, as shown by
\citet{laz03}, the polarization is marginal for $\nu>35$\,GHz. Another
possibility is the magneto--dipole emission from strongly magnetized
grains \citep{dra99}. The polarization due to this mechanism depends
on the dust composition and structure, and can be substantial at
microwave frequencies \citep{laz03}.

Finally, in the presence of a temperature quadrupole, the scattering
of the CMB photons by free electrons in galaxy clusters leads to a
secondary linear polarization. In addition to the primary CMB
quadrupole, another important origin of temperature quadrupole is due
to the peculiar velocity of clusters. However, the contribution to the
B--mode polarization from this secondary signal is considerably low,
several orders of magnitude below the lensing--induced polarization on
the large scales (for a review, see \citealt{coo04}).

\section{Galactic foregrounds subtraction and residual power spectrum}
\label{s3}

Figure \ref{f1} clearly shows that the contribution of Galactic
foregrounds to the B--mode polarization dominates over the CMB
signal at all frequencies, even in the most optimistic cases. The
knowledge of foreground emissions and the ability to remove them
assume therefore a primary role for the possibility of observing the
polarization induced by gravitational waves. Multifrequency
observations are thus absolutely necessary, expecially on large
scales. During last years, many methods have been suggested to perform
the component separation of the microwave sky. Some of them adopt
Bayesian approaches (e.g., MEM by \citealt{hob98} and Wiener filtering
by \citealt{teg96} and \citealt{bou99}
) whereas others blind techniques (e.g., SMICA by
\citealt{del03a} and FastICA by \citealt{mai02}). In this work we do
not want to go into details of different techniques and assumptions
for the components separation. Our aim is to estimate the level of
residual Galactic foregrounds left in the clean maps after their
subtraction. We make the following hypotheses: i) a template for
synchrotron and dust emission (the only Galactic foregrounds
considered) is available; ii) their spectral behaviour is estimated in
the frequency range of interest. A linear subtraction of Galactic
foregrounds is then employed extrapolating templates to the
``cosmological'' frequency. The instrumental noise present in
templates and the uncertainty on the frequency spectrum of foregrounds
are taken into account in order to compute the residual polarization.
Compared to standard techniques for component separation, this method
is, in principle, less accurate. However, it provides qualitative
estimates for the residual foregrounds that are good enough for the
aim of this work. Moreover, its simplicity allows to compute
analytically the power spectrum of residual Galactic contaminants and,
in addition, the method only requires minimal assumptions on the
foregrounds and their statistical properties.

Let $\nu_o$ be the frequency chosen to detect the CMB polarization
(hereafter called the ``observational frequency'') and $\nu_t$ the
frequency of the template for the Galactic foreground we want to
subtract. We suppose that the foreground frequency spectrum
can be approximated by a power law (at least between $\nu_t$
and $\nu_o$):
\begin{equation}
I_{\nu_o}=I_{\nu_t}\Big({\nu_o \over \nu_t}\Big)^{-\beta}\,.
\label{e5}
\end{equation}
In general, the value of the spectral index $\beta$ is expected to
change with the sky position. We indicate with $\tilde{I}_{\nu_t}
({\bf \hat{n}})$ and $\tilde\beta({\bf \hat{n}})$ the template
foreground intensity and the measured spectral index in the
sky direction ${\bf \hat{n}}$, while the difference between these
quantities and the actual ones is $\Delta I_{\nu_t}({\bf 
\hat{n}})$ and $\Delta\beta({\bf \hat{n}})$ respectively. If we
extrapolate the template to the observational frequency using Eq.
(\ref{e5}), the residual foreground left in the map at $\nu_o$ after
foreground subtraction is:
\[
\Delta I_{\nu_o}({\bf \hat{n}})=I_{\nu_o}({\bf \hat{n}})-
\tilde{I}_{\nu_o}({\bf \hat{n}})=
\]
\[
(\tilde{I}_{\nu_t}+\Delta I_{\nu_t})({\bf \hat{n}})
\Big({\nu_o \over \nu_t}\Big)^{-[\tilde\beta({\bf \hat{n}})+
\Delta\beta({\bf \hat{n}})]}-\tilde{I}_{\nu_t}({\bf \hat{n}})
\Big({\nu_o \over \nu_t}\Big)^{-\tilde\beta({\bf \hat{n}})}=
\]
\begin{equation}
\tilde{I}_{\nu_o}({\bf \hat{n}})\Big[\Big({\nu_o \over \nu_t}
\Big)^{-\Delta\beta({\bf \hat{n}})}-1\Big]+\Delta I_{\nu_t}
({\bf \hat{n}})\Big({\nu_o \over \nu_t}\Big)^{-[\tilde\beta
({\bf \hat{n}})+\Delta\beta({\bf \hat{n}})]}\,.
\label{e7a0}
\end{equation}
Keeping only first order terms, Eq. (\ref{e7a0}) is reduced to
\begin{equation}
\Delta I_{\nu_o}({\bf \hat{n}})\simeq\ln\Big({\nu_t \over \nu_o}\Big)
\tilde{I}_{\nu_o}({\bf \hat{n}})\Delta\beta({\bf \hat{n}})+
\Delta I_{\nu_t}({\bf \hat{n}})\Big({\nu_o \over \nu_t}
\Big)^{-\tilde\beta({\bf \hat{n}})}\,.
\label{e7}
\end{equation}
As expected, the residual foreground in direction ${\bf \hat{n}}$
consists in a part proportional to the uncertainty on the spectral
index, and another proportional to the error (or noise) on the
foreground polarization at the template frequency. Extending this
result to the case of polarization measurements is trivial:
\begin{eqnarray}
\Delta(Q\pm iU)_{\nu_o}({\bf \hat{n}}) & \simeq &
\ln\Big({\nu_t \over \nu_o}\Big)(\tilde{Q}\pm i\tilde{U})_{\nu_o}
({\bf \hat{n}})\Delta\beta({\bf \hat{n}})+ \nonumber \\
& & \Delta(Q\pm iU)_{\nu_t}({\bf \hat{n}})
\Big({\nu_o \over \nu_t}\Big)^{-\tilde\beta({\bf \hat{n}})}\,.
\label{e7a}
\end{eqnarray}
Assuming that no correlation exists between the errors on $\beta$ and
$Q\pm iU$ and their estimated values, the $X=E,\,B$ modes APS of the
residual polarization is the sum of the two terms in Eq. (\ref{e7a}).
The former (indicated by $C^{\mathcal{R}_1}_{X\ell}$) is obtained by
the convolution of the spectrum of the estimated foreground
polarization at $\nu_o$, $C_{X\ell}^{f}$, with that of the uncertainty
on the spectral index $\Delta\beta$, $C^{\beta}_{\ell}$. The
mathematical calculations are reported in Appendix A. We get
\begin{equation}
C^{\mathcal{R}_1}_{X\ell}={A^2 \over 16\pi}\sum_{\ell_1}
(2\ell_1+1)C^f_{X\ell_1}\mathcal{F}(\ell,\ell_1)\,,
\label{e15}
\end{equation}
where $A\equiv\ln(\nu_t/\nu_o)$ and
\[
\mathcal{F}(\ell,\ell_1)=\sum_{\ell_2}(2\ell_2+1)C^{\beta}_{\ell_2}
\Big({\ell \atop 2}{\ell_1 \atop -2}{\ell_2 \atop 0}\Big)
\Big({\ell \atop 2}{\ell_1 \atop -2}{\ell_2 \atop 0}\Big)
\]
(the last two terms are the Wigner 3j--Symbols). In principle, a
convolution is required also for the second term of Eq. (\ref{e7a}),
$C^{\mathcal{R}_2}_{X\ell}$.
However, keeping again only first order terms (in particular, we
replace $\tilde\beta({\bf \hat{n}})$ with the average value of the
spectral index, $\langle\tilde\beta\rangle$) and supposing the
polarization uncertainty in the template to behave like a white noise,
we find that
\begin{equation}
C^{\mathcal{R}_2}_{E,B\ell}=C_{\ell}^N
\Big({\nu_o \over \nu_t}\Big)^{-2\langle\tilde\beta\rangle}\,,
\label{e15b}
\end{equation}
where $C_{\ell}^N=\Omega_{pix}\sigma^2_{t}$ is the white--noise
spectrum in the template map ($\Omega_{pix}$ is the solid angle per
pixel and $\sigma_{t}$ the noise per pixel in polarization).

The hypothesis of a power--law frequency spectrum is not quite true
for the dust emission. It can be better modelled by a law
as $I_{\nu}\propto\nu^{\beta}B_{\nu}(T)$, where $B_{\nu}(T)$ is the
Planck function at temperature $T$ of dust grains.
Then, for the dust emission we replace Eq. (\ref{e5}) by
\begin{equation}
I_{\nu_o}=I_{\nu_t}\Big({\nu_o \over \nu_t}\Big)^{-\beta}
{e^{h\nu_t/KT}-1 \over e^{h\nu_o/KT}-1}\,.
\label{e5bis}
\end{equation}
We can assume the temperature of grains to be constant over all the sky
and the spatial changes in the frequency spectrum to be restricted
only to the value of $\beta$. The residual power spectra for the dust
emission will be still given by Eq. (\ref{e15}), but multiplied by the
factor $[(e^{h\nu_t/KT}-1)/(e^{h\nu_o/KT}-1)]^2$, that is $\sim2$ for
$T=18$\,K, $\nu_o=100$\,GHz and $\nu_t=140$\,GHz.

\begin{figure}
\includegraphics[width=84mm]{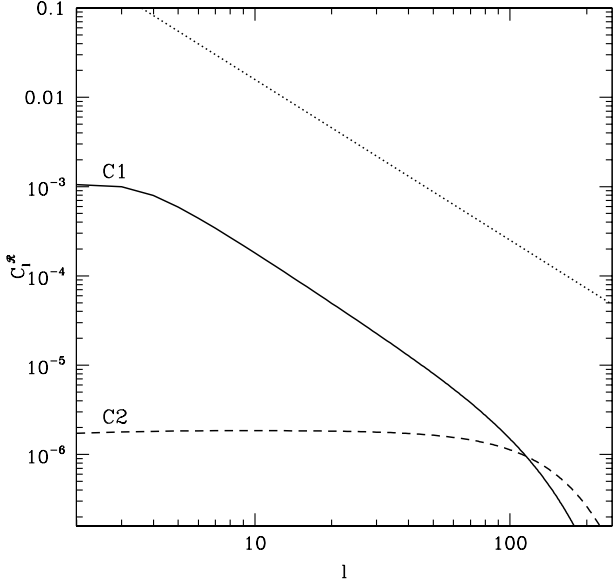}
\caption{Residual polarization spectra $C^{\mathcal{R}_1}_{X\ell}$
under the hypothesis of average (C1; solid line) and pixel--dependent
(C2; dashed line) spectral index for Galactic foregrounds. The
dotted line is a power law normalized to the unity and with slope
$-1.8$.}
\label{f2}
\end{figure}

\subsection{Residual power spectrum due to spectral index
uncertainties}

In order to estimate the power spectrum of the residual foreground
polarization by Eq. (\ref{e15})--(\ref{e15b}), the knowledge of three
angular spectra is required: $C^f_{X\ell}$ (polarization spectrum
of the Galactic foregrounds which can be directly computed from
the template maps); $C^N_{\ell}$ (spectrum of the white noise in
foreground templates); $C^{\beta}_{\ell}$ (spectrum of the spectral
index error). Here, we focus on the latter which is the most critical
to define the behaviour of $C^{\mathcal{R}_1}_{X\ell}$. We consider
two different extreme cases:

\vskip 0.2 truecm
\noindent {\bf (C1) Average spectral index}: the template for a
Galactic foreground can be extrapolated to the observational frequency
using the sky--average spectral index. In this case, the error on the
spectral index will be the difference between $\beta({\bf \hat{n}})$
and the estimated average value
$\langle\tilde\beta\rangle=\langle\beta\rangle$ (assuming no
systematic effect in the estimation). Because the spectral index of
Galactic emissions is directly related to physical conditions of the
region where the emission is produced both for synchrotron (spectral
index depends on the energy distribution of electrons) and for dust
emission (dependent on the temperature and the emissivity of grains),
we expect that maps of $\beta({\bf \hat{n}})$ and, consequently, of
the error $\Delta\beta({\bf \hat{n}})$ show coherent structures like
those observed in total--intensity templates. This is visually
observed comparing the maps of the synchrotron spectral index provided
by \citet{gia02} and \citet{ber04} and the total--intensity ones. More
quantitatively, we estimate the power spectrum of $\beta({\bf \hat{n}})$
obtained by \citet{gia02} from low--frequency surveys, and we find a
power law spectrum with a slope of $\sim-3$, in agreement with the
behaviour of the synchrotron intensity spectra (\citealt{teg96};
\citealt{bou96}; \citealt{bou99}). We expect it is same for the dust
emission.

Therefore, in this case (C1) we assume that {\bf $C^{\beta}_{\ell}$ is
proportional to the total--intensity APS for the Galactic foregrounds}.
The normalization of $C^{\beta}_{\ell}$ is obtained by fixing the
dispersion of the spectral index around its average value,
$\sigma_{\beta}$, and using the relation:
$\sigma^2_{\beta}=\sum_{\ell}(2\ell+1)C^{\beta}_{\ell}W_{\ell}/4\pi$,
where $W_{\ell}$ is the window function corresponding to the
resolution at which the dispersion $\sigma_{\beta}$ has been
computed.

The residual power spectrum $C^{\mathcal{R}_1}_{X\ell}$ obtained under
this hypothesis is shown as solid line in Figure \ref{f2}. Hereafter,
we take: $C^{\beta}_{\ell}\propto\ell^{-3}$ and $\sigma_{\beta}=0.2$,
in agreement with the results found by \citet{gia02} and \citet{ber04}
for the synchrotron emission at high Galactic latitudes, and by WMAP
measurements for the dust spectral index (see Figure 9 of
\citealt{ben03}). The polarization spectra of the Galactic foregrounds
are assumed to be of the form $C^f_{X\ell}=\ell^{-1.8}$. Both spectra
are smoothed by a window function with FWHM$=1\degr$. We notice that
the residual spectra $C^{\mathcal{R}_1}_{X\ell}$ have the same slope
of the foreground polarization spectra as expected (except for the
first $\ell$s), but with an amplitude a factor $\sim$100 lower [here
we have considered $A=1$ in Eq. (\ref{e15})].

\vskip 0.2 truecm
\noindent {\bf (C2) Pixel dependent spectral index}: as second case,
we consider that the foreground template is extrapolated to the
observational frequency using a spectral index estimated pixel by
pixel. Moreover, we suppose that the error on these spectral indeces,
$\Delta\beta({\bf \hat{n}})$, behaves like a white noise. Consequently
{\bf its power spectrum $C^{\beta}_{\ell}$ will be constant}. The
amplitude of $C^{\beta}_{\ell}$ clearly depends on the noise in data
used to compute the spectral index. As working case, we assume that
the average error on spectral index is equal to the dispersion in the
spectral index distribution, i.e. the above value $\sigma_{\beta}=0.2$.
Even if this assumption is probably very conservative, it is however
interesting because it allows us a direct comparison to results of
case C1, as the total power of the error is the same, but distributed
differently in $\ell$. Figure \ref{f2} (dashed line) shows that the
residual APS behaves like a white noise spectrum (at least up to the
resolution scale) with a very small amplitude at large angular scales:
at $\ell<10$ it is $10^{-5}$--$10^{-6}$ times lower than $C^f_{X\ell}$,
and two orders of magnitude lower than predictions for case C1.
Considering that the power of the CMB B--mode polarization is
concentrated at degree scales ($\ell<100$), this second case for
foreground subtraction is much more favourable, removing particularly
well the Galactic foregrounds on the largest angular scales.

\begin{figure*}
\includegraphics[width=84mm]{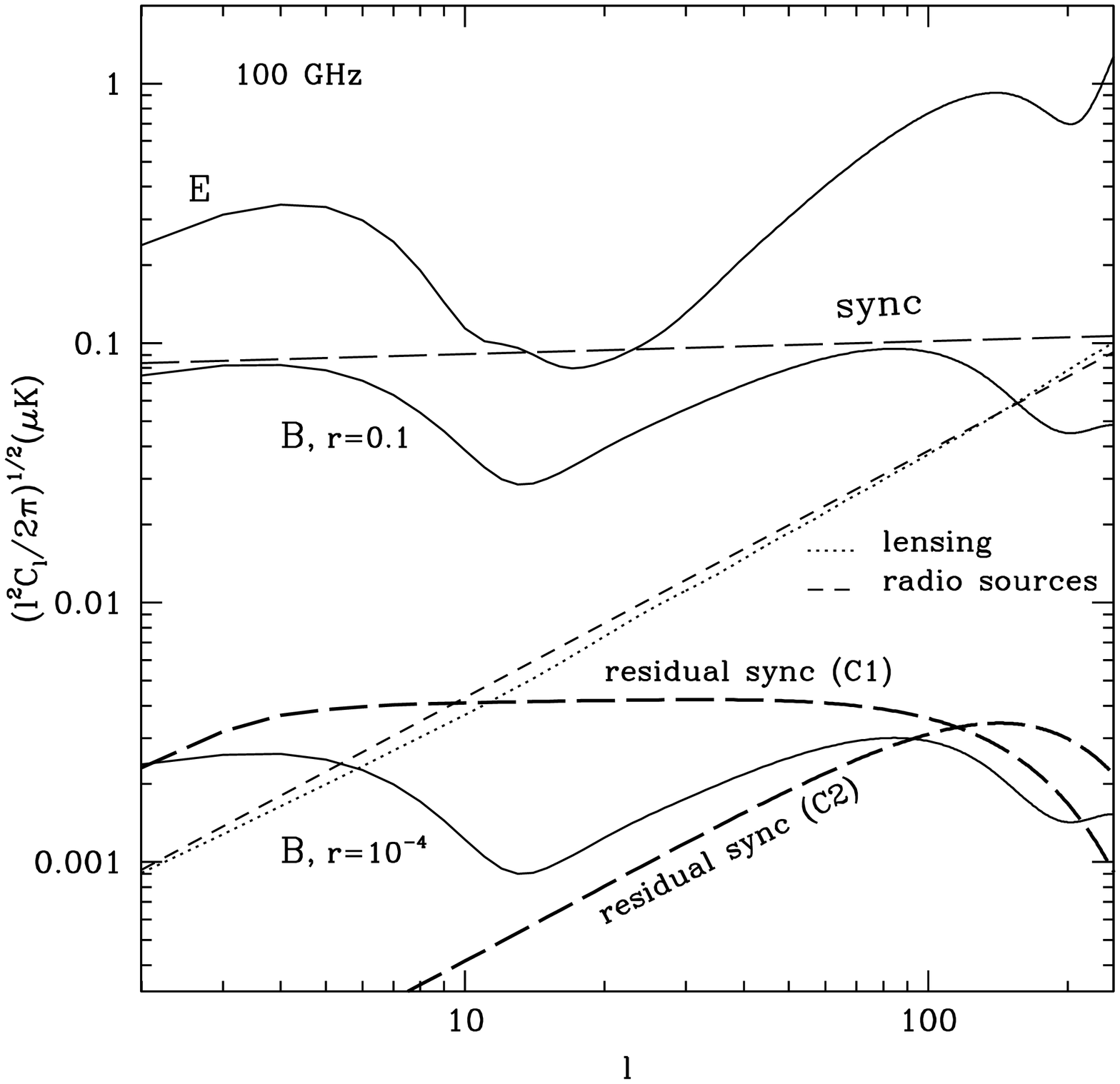}
\includegraphics[width=84mm]{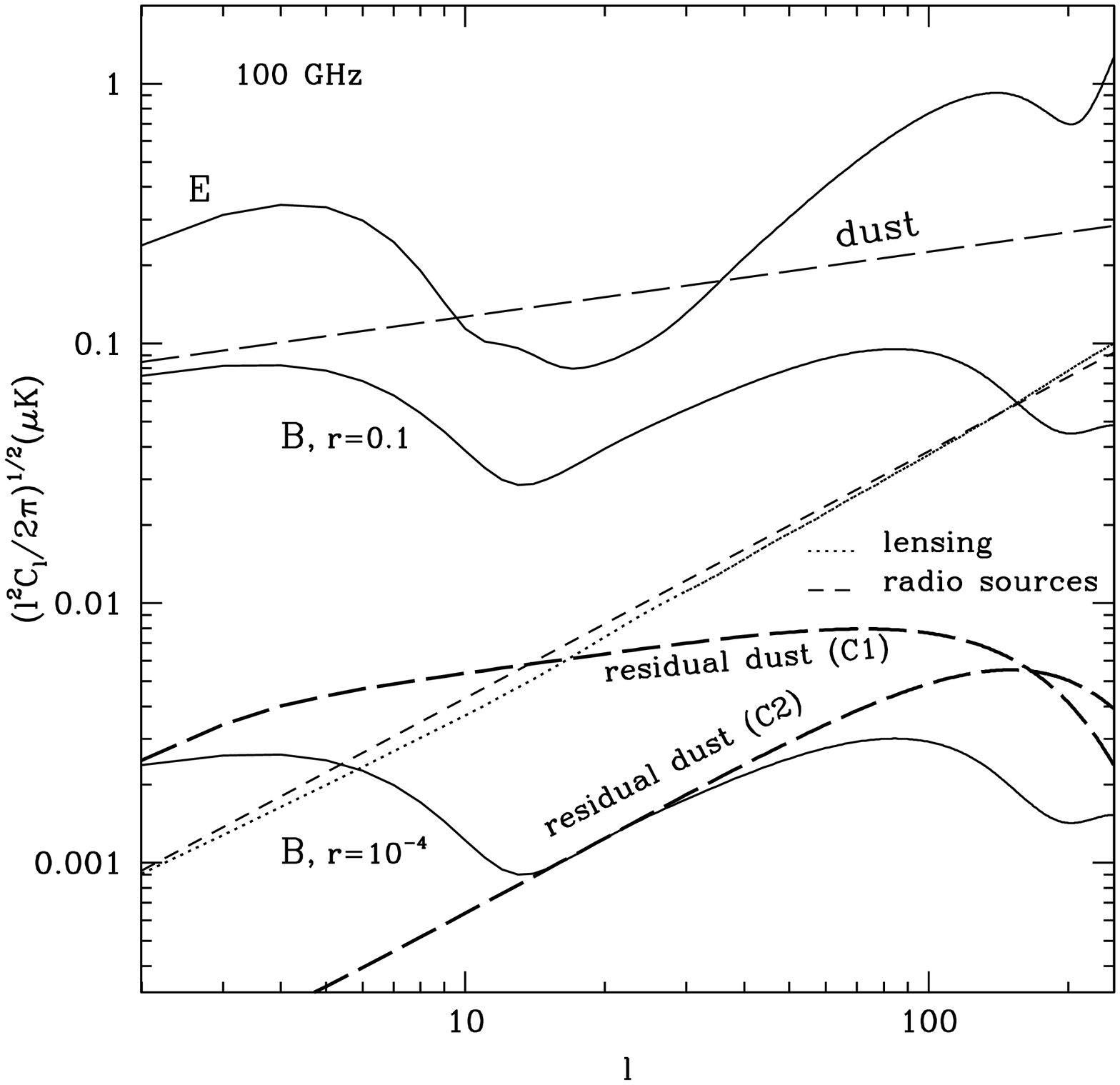}
\caption{Polarization power spectra at 100\,GHz (in thermodynamic
temperature): CMB E--mode (upper solid line) and B--mode for $r=0.1$
and $10^{-4}$ (lower solid lines); synchrotron (left plot, dashed
lines); dust (right plot, dashed lines); gravitational lensing (dotted
lines); extragalactic radio sources with $S<1$\,Jy (short dashed
lines). The thick dashed lines show the residual contribution of the
synchrotron and dust emission after their subtraction, following the
two schemes described in sec. \ref{s3}. In these spectra the effect of
the beam (FWHM$=1\degr$) start to be noticed at $\ell\sim100$.}
\label{f3}
\end{figure*}

\section{Uncertainty on the tensor--to--scalar perturbations ratio in
an ideal experiment including foregrounds}
\label{s4}

The main goal of CMB observations is to estimate the angular power
spectrum. Its amplitude and shape allows us to have information on
cosmological parameters. The precision achieved for a parameter can be
estimated from the Fisher matrix. The CMB polarization is usually
described by E and B--mode and their relative spectra; for
polarization measurements the cosmological parameter estimation using
the Fisher matrix is given by
\begin{equation}
\mathcal{F}_{ij}=\sum_{\ell}\bigg({1 \over \Delta
  C_{E\ell}^2}{\partial C_{E\ell} \over \partial\alpha_i}{\partial
  C_{E\ell} \over \partial\alpha_j}+{1 \over \Delta
  C_{B\ell}^2}{\partial C_{B\ell} \over \partial\alpha_i}{\partial
  C_{B\ell} \over \partial\alpha_j}\bigg)
\label{e3}
\end{equation}
where $\alpha_i$ are the cosmological parameters to be estimated. The
diagonal terms of the inverse of the Fisher matrix provide the minimum
possible variance for the parameter $\alpha_i$, i.e.
$(\delta\alpha_i)^2=\mathcal{F}^{-1}_{ii}$. The quantity $\Delta
C_{X\ell}$ is the uncertainty on the $X=E,\,B$ power spectra at the
multipole $\ell$:
\begin{equation}
\Delta C_{X\ell}^2={2 \over (2\ell+1)f_{sky}}(C_{X\ell}+N_{X\ell})^2\,,
\label{e4}
\end{equation}
where $f_{sky}$ is the fraction of the sky observed by the experiment.
Beside the cosmic variance, the uncertainty on $C_{X\ell}$ arises
from the noise term $N_{X\ell}$, which includes the instrumental noise
and the foregrounds contamination, treated as an extra source of
noise.

As we have previously discussed, the CMB polarization at degree scales
depends strongly on two cosmological parameters, the tensor--to--scalar
ratio $r$ and the optical depth of the Universe $\tau$. Polarization
spectra are affected also by other parameters, but polarization
measurements are of particular importance only for $r$ and $\tau$,
which can be at best poorly constrained with temperature. Therefore,
we consider in the Fisher matrix only parameters
$\alpha_{i}=\{\tau,\,r\}$, all the other cosmological parameters being
supposed to be already known. The results on the tensor--to--scalar
ratio are given in terms of the quantity $r_{lim}$, defined as the
value of $r$ under which the relative error on $r$ is 0.3 (i.e.,
corresponding to about a 3--$\sigma$ error). On the contrary, the
uncertainty on the parameter $\tau$ will be not discussed, being its
value generally estimated with a high precision, less than $10\%$, in
almost all the cases we have considered.

We consider as reference experiment a full--sky experiment with a
resolution of FWHM=$1\degr$ and a negligible level of instrumental
noise (``noise--free'').
A Galactic Plane cut is applied and $f_{sky}=0.8$ is used in the
analysis. The observational frequency, aimed to detect the
gravitational wave polarization, is choosen at $\nu=$100\,GHz, which
corresponds to the frequency where the Galactic polarized emission is
expected to be minimum (see Figure \ref{f1}). We model the
polarization of Galactic foregrounds using a power--law angular
spectrum: at 100\,GHz, in antenna temperature, the synchrotron
spectrum is
\begin{equation}
C_{X\ell}=1.2\times10^{-2}\ell^{-1.8}\,(\mu K^2)\,,
\label{e16a}
\end{equation}
giving $X_{\rm rms}=0.12\,\mu$K at $1\degr$ resolution (the
normalization is based on the WMAP results and a polarization degree
of 20$\%$ ); for the dust polarization
\begin{equation}
C_{X\ell}=1.6\times10^{-2}\ell^{-1.4}\,(\mu K^2)\,,
\label{e16b}
\end{equation}
corresponding to 0.26$\,\mu$K at $1\degr$ resolution (as expected by
the \citet{fin99} model with a polarization degree of 5$\%$). These
spectra, compared to the CMB E and B--mode ones, are plotted in Figure
\ref{f3}.

We suppose that Galactic foregrounds are subtracted from data at
100\,GHz using the prescriptions described in the previous section. In
particular, we consider that a template for the synchrotron polarized
emission is available at $\nu=60$\,GHz and one for the dust at
140\,GHz. Under these conditions and supposing noise--free templates,
the noise term in Eq. (\ref{e4}) is given only by the residual
Galactic polarization, $C^{\mathcal{R}_1}_{X\ell}$ (thick dashed lines
in Figure 4), and extragalactic foregrounds. The rms amplitude of the
Galactic polarization is reduced by a factor 20 or 60, according to
the subtraction scheme used.

\subsection{Constraints on $r$ imposed by Galactic foregrounds}
\label{s4a}

\begin{figure}
\includegraphics[width=84mm]{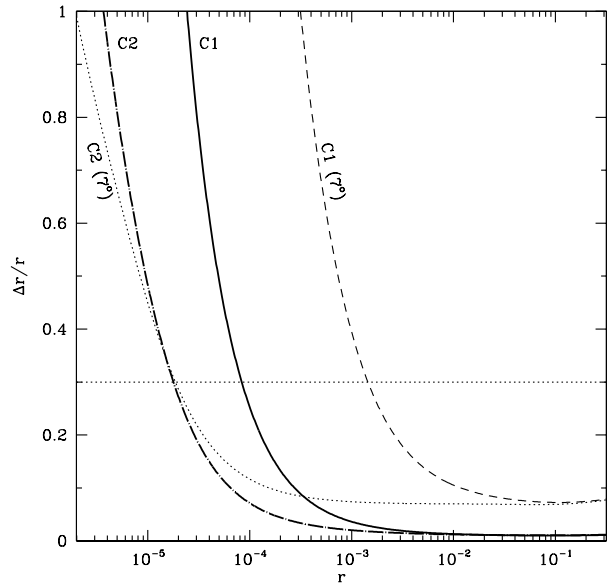}
\caption{Fractional errors of $r$ as function of $r$ for the reference
experiment (full--sky, ``noise--free'' and FWHM$=1\degr$) in the two
hypothesis on $C^{\beta}_{\ell}$: C1 (solid line) and C2 (dot--dashed
line). Dashed line (C1 case) and dotted line (C2 case) are for the
same experiment but with FWHM$=7\degr$. Only Galactic foregrounds are
considered.}
\label{f4}
\end{figure}

The Fisher matrix is used to estimate the uncertainty on $r$
for our reference experiment. No contribution from extragalactic
sources is taken into account. The results are shown in Figure
\ref{f4}. The two different methods to subtract Galactic foregrounds
are considered: in C1 case (the frequency spectrum for Galactic
emissions is considered independently of the sky position) we find
$r_{lim}=8.3\times10^{-5}$; a detection of the tensor polarization
with lower amplitude can be achieved if the spectral law is estimated
pixel by pixel (C2 case). In particular, a 3--$\sigma$ detection is
possible for $r=1.8\times10^{-5}$. These very low values for $r_{lim}$
tell us that the Galactic foregrounds are subtracted in a very
efficient way in both cases (as expected when the noise is
negligible).

No improvement of the sensitivity on $r$ is observed increasing the
instrumental resolution to angular scales smaller than $1\degr$,
confirming that all the information on the cosmological $B$--mode
comes from $\ell\la100$. If the resolution is reduced to
FWHM$=7\degr$, we find that $r_{lim}$ gets clearly worse only in the
C1 case. On the contrary, the C2 results appear to be nearly
independent of the resolution at least for FWHM$\le7\degr$, meaning
that the largest scales ($\ell\la30$) are enough to provide a good
detection of $r$. For a low--resolution experiment, the loss of
information from $\ell\sim100$ is probably balanced by a decrease of
the amplitude of $C^{\mathcal{R}_1}_{X\ell}$.

Finally, we have verified that previous results are nearly insensitive
to the value of the optical depth. In fact, the reionization affects
the B--mode APS only for $\ell\la10$, where the cosmic variance
strongly constrains the precision on the $r$ detection. Hereafter, the
results are given for $\tau=0.1$.

\subsection{Effects of the extragalactic B--mode polarization on
$r_{lim}$}
\label{s4b}

The next question is how much extragalactic radio sources and the
polarization induced by gravitational weak lensing can affect the
estimate of $r_{lim}$. At degree scales they provide only a secondary
contribution to the B--mode polarization compared to the Galactic
emissions. At 100\,GHz, the radio--sources and lensing--induced
B--mode spectra are constant with nearly the same amplitude, at least
up to $\ell\la300$. However, as shown in Figure \ref{f3},
multifrequency experiments allow us to remove efficiently the Galactic
foregrounds from polarization maps, and thus extragalactic
polarization becomes the greatest contribution to the B--mode also at
very large scales. Observations at different frequencies are not
useful for the subtraction of point sources or of the lensing--induced
polarization, that require experiments with arcminute resolution. They
become a critical problem for low--resolution experiments.

\subsubsection{Extragalactic radio sources}

In the last years, a large effort has been done to develop methods
able to detect extragalactic point sources in CMB maps and to estimate
their intensity emission (e.g., \citealt{teg98}, \citealt{san01},
\citealt{vie03}). Complete catalogues of extragalactic sources with
low flux limits are expected to be extracted from the forthcoming
data. For example, \citet{vie03}, using a method based on the
Spherical Mexican Hat Wavelet, showed that 90$\%$--completed
catalogues up to $\sim200$\,mJy can be extracted by Planck
observations at the frequencies of 100, 143 and 217\,GHz. At the
moment, these techniques have not been extended yet to polarization
maps.

In any case, because we are only interested to reduce the contribution
of point sources to the polarization maps, the best strategy is to
exploit the total--intensity catalogues in order to mask all the
pixels with bright sources. Supposing that total--intensity catalogues
of radio sources will be available up to very low flux limits, the
only restriction is given by the number of pixels that can be masked
in a map without losing too much information. This is a strong
constraint expecially for low--resolution experiments. We fix in the
ten per cent the maximum percentage of beam--size pixels we can
mask. For a full--sky experiment with FWHM$=1\degr$ (like the
reference experiment), it corresponds to 3300 pixels, taking into
account the Galactic Plane cut. According to the number counts of
\citet{tof98} model, the same number of radio sources is found at
100\,GHz for fluxes $S\ge150$\,mJy.

\begin{figure}
\includegraphics[width=84mm]{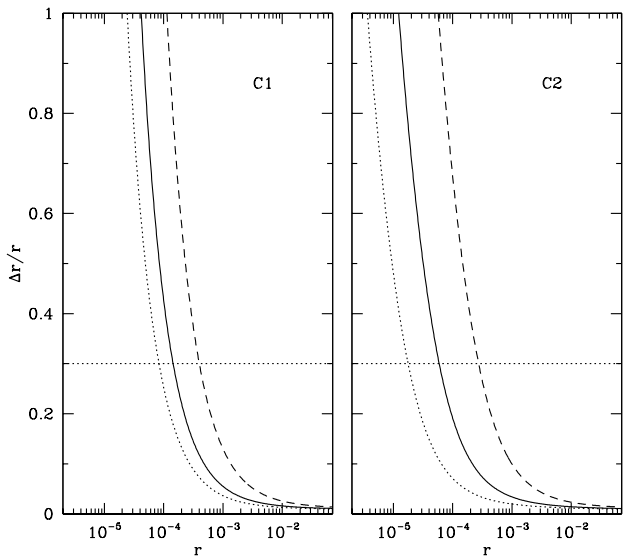}
\caption{Fractional error of $r$ for the reference experiment,
considering the two different schemes of Galactic foregrounds
subtraction (C1, left plot; C2, right plot). The contribution of radio
sources with $S<150$\,mJy (solid lines) and $S<1$Jy is included
(dashed lines). Previous results with no contribution from
extragalactic foregrounds are shown as the dotted line.}
\label{f5a}
\end{figure}

Thus, we compute the relative error on $r$ in the hypothesis that all
the sources with $S>150$\,mJy have been masked (the APS amplitude of
the remaining radio sources is $7.2\times10^{-8}\,\mu$K$^2$). We find
$r_{lim}=1.4\times10^{-4}$ and $6\times10^{-5}$ in the cases C1 and C2
respectively (see Figure \ref{f5a}). In this figure we report also
the results obtained if only the sources with $S\ge1$\,Jy are
removed. We stress that as the contribution of extragalactic sources
becomes higher the $r_{lim}$ value is less affected by the amplitude
of the residual Galactic polarization.

Because the contribution of radio sources tends to decrease at
frequencies higher than 100\,GHz, it is interesting to study what
occurs when the observational frequency is increased, $\nu_o>100$\,GHz.
We have considered $\nu_o=140$\,GHz and the Galactic foreground
templates at 60 and 220\,GHz. However, the resulting $r_{lim}$ get
significantly worse than in the reference case, confirming that
$100$\,GHz is an appropriate frequency to detect the CMB B--mode
polarization with this foreground subtraction scheme.

\begin{figure}
\includegraphics[width=84mm]{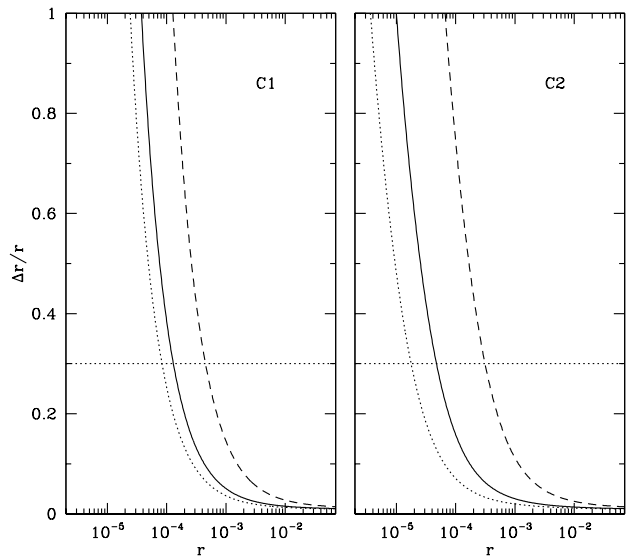}
\caption{Fractional error of $r$ for the reference experiment,
considering the two different schemes of Galactic foregrounds
subtraction (C1, left plot; C2, right plot). We include the
contribution of the lensing--induced polarization: the total one
(dashed lines) and reduced by a factor 10 (solid lines). No
contribution from extragalactic foregrounds is considered as the
dotted line.}
\label{f5b}
\end{figure}

\subsubsection{Gravitational lensing}

The decontamination of B--mode polarization maps from the
polarization induced by gravitational weak lensing can be achieved
reconstructing the projected mass distribution of large scale
structures (or the lensing potential). \citet{hu02} showed that the
best estimator for the projected mass distribution is provided by the
correlation between the CMB E and B--modes, which allows us to
reconstruct the distribution power spectrum up to multipoles of
$\ell\sim1000$. Different approaches have been suggested by
\citet{kno02}, \citet{kes02} and \citet{sel03}: they show that a
high precision in the lensing reconstruction can be obtained only for
low--noise experiments with arcminute resolution. In particular,
Table 1 of \citet{sel03} provides the residual B--mode contamination as
a function of the beam size and of the instrumental noise. The lensing
``noise'' always decreases when instrumental noise and/or the
resolution are improved. The authors suggest that in an ideal
condition of a free--noise experiment with unlimited resolution the
lensing contamination could be removed completely. On the contrary,
they argue that for wide beam detectors (FWHM$>20'$) the cleaning
methods are practically useless.

So far, we have considered that the typical polarization experiment is
full--sky with degree resolution. In this case, the lensing
contamination can become the dominant limitation for detecting the
gravitational waves polarization with low values of $r$. A 3--$\sigma$
detection, in fact, is possible only for $r\ga4\times10^{-4}$ if the
lensing noise is not subtracted (see dashed lines in Figure \ref{f5b}).
On the other hand, information from high--resolution experiments can
be exploited: the Planck mission and upcoming ground--based
experiments are expected to be able to measure the lensing--induced
B--mode APS with a good precision (\citealt{bow03}, \citealt{gan04}).
In the hypothesis that the lensing--induced spectrum can be reduced by
a factor of 10 [as indicated by \citet{kno02} for his reference
experiment], the previous results are improved significantly,
expecially in the C2 hypothesis, for which we get
$r_{lim}\simeq5\times10^{-5}$ (see Figure \ref{f5b}).

\begin{figure}
\includegraphics[width=84mm]{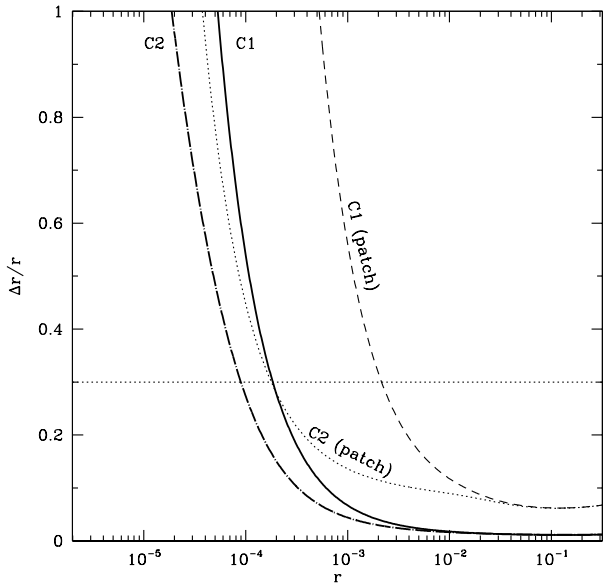}
\caption{Fractional errors of $r$ for the reference experiment,
including all foregrounds: the two schemes for the Galactic
polarization subtraction are considered (C1, solid line; C2,
dot--dashed line). Dashed line (C1) and dotted line (C2) are for a
small--area experiment without extragalactic contributions.}
\label{f7}
\end{figure}

\subsection{Limits on $r$ including all foregrounds}

Finally, we show in Figure \ref{f7} the results on $\Delta r/r$ taking
into account the following foreground contributions: 1) the residual
Galactic polarized emission; 2) the contribution from extragalactic
radio sources with flux $S<150$\,mJy; 3) the polarization induced by
weak lensing, but with its power spectrum reduced by a factor 10. For
the reference experiment we obtain $r_{lim}\simeq9\times10^{-5}$
($1.8\times10^{-4}$), using the C2 (C1) spectrum for the residual
Galactic polarization. We want to stress again that, when extragalactic
foregrounds are included, the value of $r_{lim}$ is nearly independent
of the method (and hypothesis) employed for removing the Galactic
foregrounds polarization. If we were able to completely subtract the
Galactic emission, the resulting $r_{lim}$ (under previous conditions)
would be $7\times10^{-5}$, not far from the above values. It means
that the limit of $r\sim10^{-4}$ on the detectability of the
polarization produced by gravitational waves is due essentially to
radio sources and lensing, even in the case of a raw Galactic
foregrounds subtraction (as in the C1 scheme).

When extragalactic contributions are included, low--resolution
experiments are much less effective to estimate $r$. In particular,
extragalactic radio sources are a serious problem due to the
difficulty to detect and subtract them. For example, considering
FWHM=$7\degr$, we find $r_{lim}\sim10^{-3}$ in the best case.

\subsubsection{High--resolution, small--area experiment}

An experiment with arcminute resolution could overcome the
constraints imposed by extragalactic foregrounds. In fact, on the one
hand, the number of arcminute pixels that can be masked to remove
radio sources is much higher, reducing the flux limit to $\ll150$\,mJy
and making their contribution negligible (also in a small area); on
the other hand, the high resolution (with the ``noise--free'' condition)
allows us to reconstruct with high precision the lensing B--mode power
spectrum directly from the data \citep{sel03}. In particular, we study
what occurs for an experiment covering a sky area of $30\degr\times30\degr$
in the hypothesis of a perfect subtraction of extragalactic
foregrounds: the limited sky area strongly constrains the
detectability of the tensor polarization, due to the cosmic variance
on the low $\ell$ (we remind that the uncertainty on $r$ increases
proportionally to $f_{sky}^{-1/2}$). Only if Galactic foregrounds are
removed in an accurate way, as under the C2 conditions, $r_{lim}$ is
similar to that found for a full--sky experiment (see the dotted line
in Figure \ref{f7}).

\section{Results from future experiments}

In this section we apply the previous analysis to experiments
characterized by a given sensitivity to the polarized emission.
Practically, it means to include the instrumental sensitivity in the
noise term of the Fisher matrix. We also take into account the noise
in template maps in order to estimate the APS of the residual Galactic
polarization, by adding the term given in Eq. (\ref{e15b}).

In the next years, a large number of experiments has been planned with
the explicit objective to measure the CMB polarization (for a review
on the polarization experiments see \citealt{del03b}). However, while
they may provide an accurate estimate of the E--mode power spectrum
both at large and small angular scales, for the forthcoming
experiments the detection of the gravitational waves polarization is
still a challenge and it will be possible only for high values of the
tensor--to--scalar ratio $r$.

In the mid term, the Planck
mission\footnote{http://www.astro.esa.int/SA-general/Projects/Planck/}
satisfies the best conditions to search the ``inflationary''
polarization: a full--sky coverage and a good characterization of all
the relevant foregrounds. In fact, a large frequency range is
spanned and more than one frequency is dedicated to the study of the
Galactic foregrounds, providing adequate knowledge of their spectral
behaviour and template maps in polarization. On the other hand, the
arcminutes resolution allows to detect point sources up to quite low
fluxes and to have a first estimate of the gravitational lensing
contribution. Nevertheless, the instrumental noise is still a very
strong constraint: the sensitivity to Stokes parameters $Q$ and $U$ per
square degree pixel is higher than 1\,$\mu$K at all the frequencies
(after 14 months), significantly higher than the expected rms value
for the cosmological B--mode polarization. Taking the 100\,GHz channel
as the ``observational frequency'' (and using the 30 and 217\,GHz
channels as templates for the synchrotron and dust emission
respectivelly), we find that a 2--$\sigma$ detection of the
gravitational waves polarization is possible for $r\simeq0.05$. This
result does not change if we do not include the contribution of
foregrounds.

\begin{figure}
\includegraphics[width=84mm]{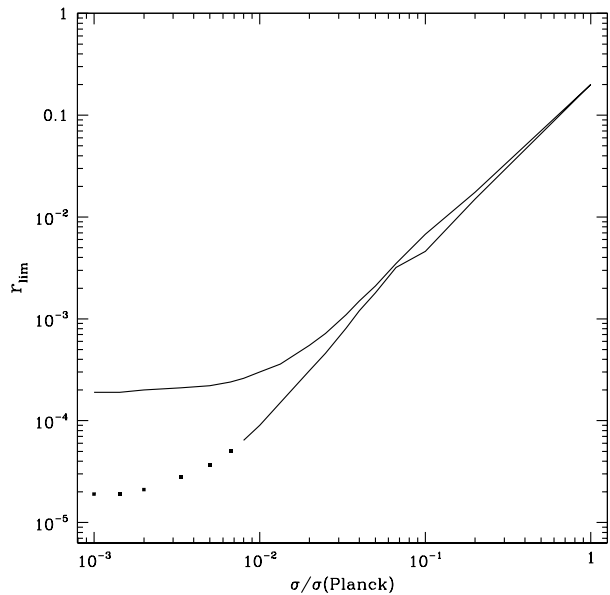}
\caption{The value of $r_{lim}$ for an experiment like Planck, but with
a sensitivity improved by a factor $\sigma/\sigma_{Planck}$. In the
upper (lower) line Galactic foregrounds are subtracted following the
C1 (C2) scheme and extragalactic foregrounds are partially
(completely) removed.}
\label{f8}
\end{figure}

The research of the gravitational waves polarization is also promising
using ground--based experiments. As discussed above, the disadvantage
of covering a limited area of the sky can be partially compensated by
the high resolution, which permits a better control of the
contribution provided by extragalactic sources and gravitational
lensing. Experiments like PolarBear (see  \citealt{gan04}), Clover
\citep{tay04} or QUIET \citep{mil04} are planned to achieve a very
good sensitivity, of the order of $0.1\,\mu$K or less, using arrays of
1000s of detectors in only two or three frequency channels. As
example, we discuss in detail the case of PolarBear: it will
operate at the frequencies 90, 150 and 220\,GHz covering 225 square
degrees; the number of detectors for each frequency should be 400, 600
and 200 with a sensitivity per detector of 310, 345 and
640\,$\mu$Ks$^{1/2}$ respectively. Considering an observational time
of 1.5 years with an efficiency of 30$\%$, the sensitivity per
$1\degr$ pixel at 90\,GHz would be $\simeq0.1\,\mu$K, an order of
magnitude lower than the Planck one. We find that PolarBear will be
able to provide a 2--$\sigma$ detection of the tensor polarization for
$r\simeq0.07$.

Based on the above experiments, we describe below the possible
features for future missions specifically dedicated to the detection
of the CMB B--mode polarization, and we compute the corresponding
detectability limit on $r$.

\vskip 0.2truecm
\noindent {\bf Space experiment}: 
it is interesting to know how the detectability limit on $r$ improves
when reducing the instrumental noise in the Planck experiment by
the same factor at all the frequencies. Figure \ref{f8} shows the
results for $r_{lim}$. In the upper line we are supposing that
Galactic foregrounds are subtracted following the C1 scheme and
extragalactic foregrounds are partially reduced (as discussed in
Sec. 4.3). This is probably an upper limit. On the other hand, in the
lower line extragalactic foregrounds are neglected (we consider they
are completely removed), while the C2 scheme is used for the Galactic
polarization subtraction. We observe that the results are independent
of the two hypothesis on the foregrounds subtraction if
Planck--detectors sensitivity is improved by a factor $\la30$
(corresponding to $r_{lim}\ga10^{-3}$). In this case, the instrumental
noise is still the most constraining contribution. For sensitivities 100
times or more better than the Planck one, the upper line is constant
with $r_{lim}\simeq3.5\times10^{-4}$, while the lower line still
decreases reaching a $r_{lim}$ similar to the free--noise case for
$\sigma/\sigma_{Planck}\sim10^{-3}$. We plot the tail of the lower
curve as a dotted line in order to indicate that, for so high
sensitivities, it must not be considered a lower limit for the
detectability of $r$, since the uncertainty of foregrounds spectral
index we used, $\sigma_{\beta}=0.2$, can be improved.

We discuss now a more concrete example for a space experiment: a
total of 10,000 detectors in three channels at 100, 140 and 220\,GHz
with the same sensitivity of Planck detectors. With about 3,000
detectors for each channel, instead of 8 present in Planck, we expect
the sensitivity to increase by a factor of about 20. The 100\,GHz
channel is used as observational frequency and the other two channels
to remove the dust polarized contribution. A template for the
synchrotron polarization is also required, being its signal still
important at 100\,GHz. If the low--frequency channels of the Planck
mission are considered, we find
$r_{lim}\simeq2.5\times10^{-3}$. In spite of the low accuracy in
the synchrotron subtraction, $r_{lim}$ is only a factor $\sim1.4$
higher than the corresponding value of Figure \ref{f8}. The results
are not affected by the contribution of extragalactic foregrounds
(radio sources and lensing), while only a slight dependence is noticed
on the scheme of the Galactic foregrounds subtraction.

\vskip 0.2truecm
\noindent {\bf Ground Based Bolometers}: an array of 5,000 bolometers
at the frequencies 90 and 150\,GHz with a sensitivity per detector of
250\,$\mu$Ks$^{1/2}$. After one year of observations (60$\%$
efficiency) over a $30\degr\times30\degr$ patch of the sky, we obtain
a sensitivity of 33\,nK per 1--degree pixel at both frequencies. Using
the 30--GHz channel\footnote{In this case, the residual polarization
arises mostly from the noise in the template map, rather than from the
uncertainty on the spectral index. So, foreground templates at frequencies
distant from the observational one are preferred.} of Planck mission
to subtract the synchrotron emission, we find $r_{lim}=0.03$.
These values can be reduced by a factor 2 if a channel at 100\,GHz is
choosen instead of 90\,GHz. The results are practically
independent of the scheme to subtract Galactic foregrounds and of the
contribution of extragalactic foregrounds. Reducing the area observed,
the sensitivity per pixel is improved but not the sensitivity on $r$:
for a $20\degr\times20\degr$ area, $\sigma=22$\,nK while
$r_{lim}\simeq0.07$.

\vskip 0.2truecm
\noindent {\bf Ground Based Radiometers}: an array of 1000 radiometers
at 30\,GHz and 4000 at 90\,GHz. Assuming 1 year of observations over a
$30\degr\times30\degr$ patch with a 80$\%$ efficiency and a
sensitivity of 120 and 200\,$\mu$Ks$^{1/2}$ at 30 and 90\,GHz
respectivelly, we get sensitivities per square degree pixel of
31.5\,nK at 30\,GHz and 21\,nK at 90\,GHz. Using the 217--GHz channel$^5$
of Planck for the subtraction of the dust emission, we find a
2--$\sigma$ detection for $r\simeq0.28$. This result points out that
experiments based only on radiometers, although with comparable or
better sensitivity, appear to be less efficient than bolometers to
detect the tensor polarization [or, in other terms, that higher
frequencies ($\nu\ga100$\,GHz) are better than lower ones
($\nu\la100$\,GHz)]. The reason is the different spectral behaviour
between dust and synchrotron emission: the frequency spectrum of
synchrotron is very steep and already at 100\,GHz its polarized
contribution is small; on the contrary, the dust spectrum decreases
quite slowly with the wavelength and at 70--80\,GHz its polarization
level is still comparable to the synchrotron one.

Finally, considering jointly radiometers that operate at 30
and 70\,GHz and bolometers at 100 and 150\,GHz with the features
previously described, we find a sensitivity to the CMB B--mode
polarization similar to the one achieved by the space mission:
$r_{lim}\sim6\times10^{-3}$.

\section{Discussions and Conclusions}

In this paper we have investigated the constraints that foregrounds
can impose on the detection of the B--mode polarization induced by
gravitational waves. First of all, we carry out a detailed analysis on
the expected level of the foregrounds polarization at degree scales,
which are the angular scales where the cosmological B--mode signal
is concentrated. 
The diffuse Galactic emissions (synchrotron and dust emission) are
the dominant components. The minimum of their polarized emission is
found to be around 100\,GHz, corresponding to a rms value for the
B--mode of $\sim0.5\,\mu$K at $1\degr$ resolution (to be compared with
the CMB $B_{\rm rms}\simeq0.43\sqrt{r}\,\mu$K). We expect a shift of
the ``cosmological window'' in polarization towards higher frequencies
as compared to temperature fluctuations, where the lowest emissivity
of foregrounds is observed around 60\,GHz. It is due to the high
polarization degree of the synchrotron emission, whose diffuse
component is observed to be up to 30$\%$ polarized or more. The
contribution of extragalactic radio sources and of the polarization
produced by gravitational lensing is secondary on large scales, but
not negligible when techniques to subtract Galactic foregrounds are
applied to polarization data. On the contrary, other sources of
polarized emission (such as infrared galaxies, SZ effect, etc.) are
expected to be negligible.

The subtraction of the Galactic contributions from polarization maps
is mandatory to search for the gravitational waves signal.
Assuming component separation boils down to subtracting, at the CMB
observing frequency, a set of foreground templates scaled in frequency
according to a model emission law,
we provide an analytical expression for the power
spectrum of the residual polarization, depending on the noise in the
foreground templates and on the uncertainty in their spectral
behaviour.
We study two concrete cases: i-- the frequency spectrum of Galactic
emissions is supposed independent of the sky direction; ii-- the frequency
spectrum is estimated pixel by pixel, but with an uncertainty that
behaves like white noise. As expected, under this second hypothesis a
better subtraction of Galactic foregrounds is achieved: if only the
contribution of the residual
Galactic foregrounds is taken into account, we find that a 3--$\sigma$
detection of the tensor polarization is possible for $r\sim10^{-5}$,
corresponding to an energy of inflation of $\sim2\times10^{15}$\,GeV.
However, this value should not be considered as a lower limit for the
detectability of the primordial B--mode polarization since it arises
from working hypothesis. In fact, in this case, the amplitude of the
residual Galactic APS depends on the average uncertainty with which
the spectral index is computed. We have taken $\sigma_{\beta}=0.2$:
this value is probably reliable for the data available up to now, but
it will be improved in the future. In this sense, a limit on the
detectability of the CMB B--mode polarization cannot be put ``a
priori''. Opposite conclusions are found if the subtraction of
Galactic foregrounds is based on the assumption that their spectral
law is independent of the sky direction. Now, the amplitude of the
residual Galactic foregrounds depends on the intrinsic dispersion of
the spectral law in the sky ($\sigma_{\beta}=0.2$) and not on our
ability to estimate it. So, the value of $r\simeq10^{-4}$ that we find
for a 3--$\sigma$ detection corresponds to the lower limit for the
detectability of the gravitational waves polarizationin this case.

Clearly then, taking into account
the spectral variability of foregrounds in the component separation
is a key point in order to detect cosmological CMB B--mode
polarization for very low $r$. At the present time, only few methods
have been proposed to perform the component separation with spatial
variability of foregrounds \citep{sto04,ben03} showing promising
results. More efforts within this field are still required.

The effects of extragalactic radio sources and gravitational
lensing on the $r$ sensitivity is also studied. After subtraction of
the Galactic foregrounds, they become the major contaminant also on
the large scales. In principle, CMB experiments with high resolution
can control accurately these foregrounds. Lensing and radio sources
are much more problematic for measurements with a wide beam and
complementary data with high resolution are required to reduce their
polarized contribution. We have seen that, for a full--sky
``noise--free'' experiment with $1\degr$ resolution, a reduction by a
factor 10 for the B--mode APS of radio sources and lensing is probably
feasible, giving a detectability limit of $r\sim10^{-4}$ independent
of the scheme to subtract Galactic foregrounds. This value is, in any
case, better than those found by experiments covering only a small area
of the sky, that are strongly limited by the cosmic variance.

Finally, we have discussed what are the prospects, ``in practice'',
for the detection of tensor polarization in the future. In the near
term, a tensor--to--scalar ratio of $r\sim0.05$ is well within reach
of the Planck mission and planned ground--based experiments. A
detection of tensor modes with so high values of $r$ would be an
indication of an inflation driven by some exotic physics \citep{kin03}.
The next generation of missions, that are at the moment under
discussion by NASA and ESA, will be able to significantly improve the
sensitivity on the polarized emission by means of arrays of
$10^3$--$10^4$ detectors, and will probably allow us to investigate
more conventional theories of the inflation, spanning ranges of energy
down to $\sim5\times10^{15}$GeV.

\vskip 0.7truecm \noindent {\it Acknowledgements}. MT and EMG gratefully
acknowledges the financial support provided through the European
Community's Human Potential Programme under contract
HPRN-CT-2000-00124, CMBNET, and through the Spanish Ministerio de
Ciencia y Tecnologia, reference ESP2002-04141-C03-01. EMG, PV and JD
kindly thank financial support from a joint Spain-France project
(Acci\'on Integrada HF03-163, Programmes d'Actions
Int\'egr\'ees (Picasso)).

\onecolumn
\appendix
\section{Estimate of the residual power spectrum}

In this appendix we demonstrate the result given in Eq. (\ref{e15}).
In particular, we want to calculate the E-- and B--mode power spectrum
for the function
$\mathcal{R}({\bf \hat{n}})=\mathcal{P}({\bf \hat{n}})\beta({\bf \hat{n}})$,
where $\mathcal{P}$ is a spin--2 function [in Eq. (\ref{e15}) it is
the linear combination of Stokes parameters $Q\pm iU$] and $\beta$ is
a real scalar function (i.e., the error on the spectral index).

A spin--2 function can be expanded on the sphere by the spin--2
spherical harmonics: $f({\bf \hat{n}})=\sum_{\ell m}a_{2,\ell m ~2}
Y_{\ell m}({\bf \hat{n}})$ and $f^*({\bf \hat{n}})=\sum_{\ell m}
a_{-2,\ell m ~-2}Y_{\ell m}({\bf \hat{n}})$. According to
\citet{zal97} formalism, the $B$--mode power spectrum is
\begin{eqnarray}
C_{B\ell}=<|a_{B\ell m}|^2> & = & {1 \over 4}<(a_{-2,\ell m}-
a_{2,\ell m})(a^{*}_{-2,\ell m}-a^{*}_{2,\ell m})>= \nonumber \\
& = & {1 \over 4}\Big(<a_{2,\ell m}a^{*}_{2,\ell m}>+
<a_{-2,\ell m}a^{*}_{-2,\ell m}>-<a_{-2,\ell m}a^{*}_{2,\ell m}>-
<a_{2,\ell m}a^{*}_{-2,\ell m}>\Big)\,.
\label{e8}
\end{eqnarray}
In order to compute $C^{\mathcal{R}}_{B\ell}$, we start by considering
the first term in the right hand of Eq. (\ref{e8})
\begin{equation}
<a^{\mathcal{R}}_{2,\ell m}a^{\mathcal{R}*}_{2,\ell m}>=
<\int d\Omega\,_2Y^*_{\ell m}({\bf \hat{n}})\mathcal{P}({\bf \hat{n}})
\beta({\bf \hat{n}})\int d\Omega'\,_{-2}Y^*_{\ell -m}({\bf \hat{n}'})
\mathcal{P}^*({\bf \hat{n}'})\beta({\bf \hat{n}'})>\,.
\label{e9}
\end{equation}
Using the appropiate expansions for
$\mathcal{P}({\bf \hat{n}})=\sum_{\ell m}a^{\mathcal{P}}_{2,\ell m ~2}
Y_{\ell m}({\bf \hat{n}})$ and $\beta({\bf \hat{n}})=
\sum_{\ell m}a^{\beta}_{\ell m}Y_{\ell m}({\bf \hat{n}})$, and removing
from integrals all the terms independent of the sky position, we find
\begin{eqnarray}
<a^{\mathcal{R}}_{2,\ell m}a^{\mathcal{R}*}_{2,\ell m}> & = &
<\sum_{\ell_1m_1}\sum_{\ell_2m_2}a^{\mathcal{P}}_{2,\ell_1m_1}
a^{\beta}_{\ell_2m_2}
\int d\Omega\,_2Y^*_{\ell m}({\bf \hat{n}})\,_2Y_{\ell_1m_1}({\bf \hat{n}})
Y_{\ell_2m_2}({\bf \hat{n}})\sum_{\ell_3m_3}\sum_{\ell_4m_4}
a^{\mathcal{P}}_{-2,\ell_3m_3}a^{\beta}_{\ell_4m_4}\times \nonumber \\
& \times & \int d\Omega'\,
_{-2}Y^*_{\ell-m}({\bf \hat{n}'})\,_{-2}Y_{\ell_3m_3}
({\bf \hat{n}'})Y_{\ell_4m_4}({\bf \hat{n}'})>\,=\,
\sum_{\ell_1m_1}\sum_{\ell_2m_2}\sum_{\ell_3m_3}
\sum_{\ell_4m_4}<a^{\mathcal{P}}_{2,\ell_1m_1}a^{\mathcal{P}}_{-2,\ell_3m_3}>
<a^{\beta}_{\ell_2m_2}a^{\beta}_{\ell_4m_4}>\times \nonumber \\
& \times & \int d\Omega\,_2Y^*_{\ell m}({\bf \hat{n}})\,
_2Y_{\ell_1m_1}({\bf \hat{n}})Y_{\ell_2m_2}({\bf \hat{n}})
\int d\Omega'\,_{-2}Y^*_{\ell-m}({\bf \hat{n}'})\,_{-2}Y_{\ell_3m_3}
({\bf \hat{n}'})Y_{\ell_4m_4}({\bf \hat{n}'})\,.
\label{e10}
\end{eqnarray}
Now, taking into account that
$_2Y_{\ell m}=\,_{-2}Y^*_{\ell-m}$\,,\,
$<a^{\mathcal{P}}_{2,\ell_1m_1}a^{\mathcal{P}}_{-2,\ell_3m_3}>=
(-)^{m_1}<|a^{\mathcal{P}}_{2,\ell_1m_1}|^2>
\delta_{\ell_1\ell_3}\delta_{m_1-m_3}$ and $<a^{\beta}_{\ell_2m_2}
a^{\beta}_{\ell_4m_4}>=(-)^{m_2}C^{\beta}_{\ell_2}
\delta_{\ell_2\ell_4}\delta_{m_2-m_4}$, we obtain
\begin{eqnarray}
<a^{\mathcal{R}}_{2,\ell m}a^{\mathcal{R}*}_{2,\ell m}> & = &
\sum_{\ell_1\ell_2}<|a^{\mathcal{P}}_{2,\ell_1m_1}|^2>
\,C^{\beta}_{\ell_2}\times \nonumber \\
& & \times\sum_{m_1m_2}(-)^{m_1+m_2}\int d\Omega\,_{-2}
Y_{\ell-m}({\bf \hat{n}})\,
_2Y_{\ell_1m_1}({\bf \hat{n}})Y_{\ell_2m_2}({\bf \hat{n}})
\int d\Omega'\,_2Y_{\ell m}({\bf \hat{n}'})\,_{-2}Y_{\ell_1-m_1}
({\bf \hat{n}'})Y_{\ell_2-m_2}({\bf \hat{n}'})\,.
\label{e11}
\end{eqnarray}
According to properties of spin--weighted spherical harmonics, we have
\begin{equation}
\int d\Omega\,_{s_1}Y_{\ell_1m_1}({\bf \hat{n}})\,_{s_2}
Y_{\ell_2m_2}({\bf \hat{n}})\,_{s_3}Y_{\ell_3m_3}({\bf \hat{n}})=
{(-)^{s_1+m_1} \over \sqrt{4\pi}}\Big[\Pi_{i=1}^3(2\ell_i+1)\Big]^{1/2}
\Big({\ell_1 \atop -s_1}{\ell_2 \atop -s_2}{\ell_3 \atop -s_3}\Big)
\Big({\ell_1 \atop m_1}{\ell_2 \atop m_2}{\ell_3 \atop m_3}\Big)\,.
\label{e12}
\end{equation}
The last two terms are the Wigner 3j--Symbols and verify the
condition
\begin{displaymath}
\Big({\ell_1 \atop m_1}{\ell_2 \atop m_2}{\ell_3 \atop m_3}\Big)\ne0
~~~~~~{\rm if}~~\left\{ \begin{array}{l}
m_1+m_2+m_3=0 \\
|\ell_2-\ell_3|\le\ell_1\le|\ell_2+\ell_3| \end{array}\right.
\label{e13a}
\end{displaymath}
and the orthogonality relation
\begin{equation}
\sum_{m_2m_3}(2\ell_1+1)
\Big({\ell_1 \atop m_1}{\ell_2 \atop m_2}{\ell_3 \atop m_3}\Big)
\Big({\ell_1 \atop m_1}{\ell_2 \atop m_2}{\ell_3 \atop m_3}\Big)=1\,.
\label{e13b}
\end{equation}
Hence, Eq. (\ref{e11}) can be simplified
\begin{eqnarray}
<a^{\mathcal{R}}_{2,\ell m}a^{\mathcal{R}*}_{2,\ell m}> & = &
\sum_{\ell_1\ell_2}<|a^{\mathcal{P}}_{2,\ell_1m_1}|^2>C^{\beta}_{\ell_2}
\sum_{m_1m_2}(-)^{m_1+m_2}{(2\ell+1)(2\ell_1+1)(2\ell_2+1) \over 4\pi}
\times \nonumber \\
& & \times\Big({\ell \atop 2}{\ell_1 \atop -2}{\ell_2 \atop 0}\Big)
\Big({\ell \atop -m}{\ell_1 \atop m_1}{\ell_2 \atop m_2}\Big)
\Big({\ell \atop -2}{\ell_1 \atop 2}{\ell_2 \atop 0}\Big)
\Big({\ell \atop m}{\ell_1 \atop -m_1}{\ell_2 \atop -m_2}\Big)
\nonumber \\
& = & {1 \over 4\pi}\sum_{\ell_1}(2\ell_1+1)
<|a^{\mathcal{P}}_{2,\ell_1m_1}|^2>\sum_{\ell_2}(2\ell_2+1)
C^{\beta}_{\ell_2}
\Big({\ell \atop 2}{\ell_1 \atop -2}{\ell_2 \atop 0}\Big)
\Big({\ell \atop 2}{\ell_1 \atop -2}{\ell_2 \atop 0}\Big)\,.
\label{e14}
\end{eqnarray}
It is easy to verify that an equivalent result is found for all the
other terms in Eq. (\ref{e8}). Finally, we get:
\begin{equation}
C^{\mathcal{R}}_{B\ell}={1 \over 16\pi}\sum_{\ell_1}(2\ell_1+1)
C^{\mathcal{P}}_{B\ell_1}\sum_{\ell_2}(2\ell_2+1)C^{\beta}_{\ell_2}
\Big({\ell \atop 2}{\ell_1 \atop -2}{\ell_2 \atop 0}\Big)
\Big({\ell \atop 2}{\ell_1 \atop -2}{\ell_2 \atop 0}\Big)\,.
\label{e15app}
\end{equation}
We remind that the sums on $\ell_1$ and $\ell_2$ are done only for
values $|\ell_1-\ell_2|\le\ell\le\ell_1+\ell_2$. The same result can
be found for the E--mode spectrum.

\end{document}